# Photocurrent Nanoscopy of Quantum Hall Bulk

Ran Jing[1*], Boyi Zhou[2,3], Jiacheng Sun[2], Shoujing Chen[2], Wenjun Zheng[2], Zijian Zhou[2], Heng Wang[2], Lukas Wehmeier[2,5], Bing Cheng[2], Michael Dapolito[2,4], Yinan Dong[4], Zengyi Du[2], G. L. Carr[5], Xu Du[2], D. N. Basov[4], Qiang Li[1,2], Mengkun Liu[2,5*]

[1]*Condensed Matter Physics and Materials Science Division, Brookhaven National Laboratory; Upton, NY, 11973-5000, USA.*

[2] *Department of Physics and Astronomy, Stony Brook University; Stony Brook, New York 11794, USA.*

[3]*Center for Integrated Science and Engineering, Columbia University, New York, New York 10027, USA*

[4]*Department of Physics, Columbia University; New York, New York 10027, USA.*

[5]*National Synchrotron Light Source II, Brookhaven National Laboratory; Upton, New York 11973, USA.*

*rjing@bnl.gov

*mengkun.liu@stonybrook.edu

**Abstract:**

Understanding nanoscale electronic and thermal transport of two-dimensional (2D) electron systems in the quantum Hall regime, particularly in the bulk insulating state, poses considerable challenges. One of the primary difficulties arises from the presence of chiral edge channels, whose transport behavior obscures the investigation of the insulating bulk. Using near-field (NF) optical and photocurrent (PC) nanoscopy, we probe real-space variations of the optical and thermal dynamics of graphene in the quantum Hall regime without relying on complex sample or electrode geometries. Near the charge neutrality point (CNP), we detect strong optical and photothermal signals from resonant inter-Landau level (LL) magnetoexciton excitations between the 0th and ±1st LLs, which gradually weaken with increasing doping due to Pauli blocking. Interestingly, at higher doping levels and full integer LL fillings, photothermal signals reappear across the entire sample over a ~10-micrometer scale, indicating unexpectedly long cooling lengths and nonlocal photothermal heating through the insulating bulk. This observation suggests thermal conductivity persists for the localized states even as electronic transport is suppressed – a clear violation of the Wiedemann-Franz (WF) law. Our experiments provide novel insights into nanoscale thermal and electronic transport in incompressible 2D gases, highlighting the roles of magnetoexcitons and chiral edge states in the thermo-optoelectric dynamics of Dirac quantum Hall state.

**Introduction**

The study of electric, thermal, and optical properties of the two-dimensional (2D) electron gas under magnetic fields is pivotal for advancing our understanding of topological quantum states with the breaking of time-reversal symmetry. In the quantum Hall regime, the Hall resistance of 2D systems is quantized with remarkable precision, largely independent of details of samples due to the topological nature of edge-dominated transport [1, 2, 3, 4, 5, 6]. On the other hand, the electric conductivity ($\overleftrightarrow{\sigma(\omega)}$), thermoelectric coefficient ($\overleftrightarrow{\alpha}$), and thermal conductivity ($\overleftrightarrow{\kappa}$) tensor exhibit dramatic changes and spatial fluctuations in response to electric gating, magnetic fields, or defects, revealing unique quantum behaviors that pervade the entire sample [7, 8, 9]. However, studying the bulk (non-edge) properties of a quantum Hall 2D electron gas presents significant challenges. For instance, traditional transport measurements often fail to capture the nuanced behavior of the bulk because they average responses over large spatial areas. To access the bulk heat or electric transport without interference from edge contributions, specially designed circuits are often necessary [10, 11, 12]. Moreover, imaging techniques face difficulties in capturing electrical, thermal, and optical properties simultaneously, which are intimately correlated and vital to distinguishing the material's behavior at the microscopic level [13, 14, 15, 16, 17, 18, 19].

In this paper, we explore the nanoscale photothermal and optical properties of graphene edge and bulk in the quantum Hall regime. Due to its linear dispersion and high carrier mobility, graphene exhibits robust quantum Hall effects at relatively higher temperatures compared to conventional 2D electron gas [4, 5]. In semiconductor systems [10], a mild violation of Wiedemann-Franz (WF) law is observed in the localized state of quantum Hall bulk. The violation is believed to arise from the Coulomb interaction between edge states bounded on the widespread micro charge puddles in the bulk. In graphene, theories [20, 21, 22] predicted a much stronger violation of WF law. The violation can be produced directly by calculating the thermal Kubo formula with Dirac band structure and a finite impurity scattering. At finite temperatures, the thermal conductivity of graphene can exhibit an 'anti-phase' quantum oscillation compared to the electrical conductivity. Despite theoretical predictions [20, 21, 22] and hints in other material systems [10, 23], experimental evidence for the violation of WF law in graphene is still lacking.

By mapping tip-induced photocurrent in graphene at the nanoscale (**Fig.1a**), we examine how the photothermal behavior varies on and off the quantized Landau levels (LLs). The tip-enhanced electromagnetic field leads to a local photo-absorption and current generation in graphene at the tip apex due to photothermal effect. When the sample is raster-scanned under the tip, the photo-induced thermoelectric currents can be recorded by electrodes and mapped at various sample locations, eliminating the need for engineering complicated thermal sources. In the localized state where bulk current is strongly suppressed, carrier diffusion supported by the edge state becomes a convenient teller of the tip-initiated thermal transport through the bulk. This capability to probe bulk thermal transport without introducing new edges (e.g. through electrodes) represents a critical advantage when studying insulating bulk sample regions, where conventional understanding suggests that both DC electric and thermal currents should be prohibited.

The local carrier doping levels at the tip apex are calibrated using the near-field optical responses with a spatial resolution of ~20 nm (**see methods**). At cryogenic temperatures in the integer quantum Hall regime, we observe unexpectedly long-range photothermal transport in the bulk of the sample. At high Landau levels, this long cooling length appears at full filling of the Landau levels (QH plateaus), likely induced by an anomalously high in-plane thermal conductivity and a low cooling rate of electron to phonon system. This high thermal conductivity persists at the quantum Hall plateaus, supporting a strong violation of the WF law in graphene at finite temperature [20, 21, 22]. At low Landau levels, the long cooling length we found can be attributed to the light-induced magnetoexciton polariton propagation, which effectively transports energy from the sample's center to its edge [24]. Areal scanning at these specific Landau fillings reveals unique edge-bulk correspondence of a pristine sample in the clean limit.

The monolayer graphene (MLG) used in our study is encapsulated in two 10 nm hexagonal boron nitride (hBN) layers. The sample was configured with a back gate applied through dielectric layers composed of 10 nm hBN and 285 nm $SiO_2$. Given that each LL in MLG exhibits a four-fold degeneracy, we use the filling factor $\nu$ to conveniently label the doping levels. For instance, the $0^{th}$ LL spans the range of $-2<\nu<2$, while the $1^{st}$ LL extends from $2<\nu<6$, and so forth. At $\nu=6$ and $\nu=10$, graphene has the $1^{st}$ and $2^{nd}$ LL fully filled, respectively, causing the bulk of the sample to enter localized states (**Fig. 1b**). The $0^{th}$ LL of hBN encapsulated graphene is usually split with a gap $2\Delta$ due to the lifted valley degeneracy [25, 26, 27, 28, 29, 30, 31]. The cyclotron energy of CNP graphene, corresponding to the LL transition between the $\pm 0^{th}$ and the $\pm 1^{st}$ LL, is therefore defined by $\omega_{c\pm} = \sqrt{2eBv_F^2 + \Delta^2} \pm \Delta$. In addition, we also introduce a controlled change of doping and substrate symmetry via patterning a particular region of the $SiO_2$ layer into a superlattice structure (as enclosed by the black dashed lines in the first panel in **Fig. 1c**). We index the regions inside and outside the patterned substrate as region 1 (superlattice substrate) and 2 (pristine graphene).

For our nanoscopy experiments, we harness the optical illumination from a QCL laser (Hedgehog from Daylight Solution) which emits CW light at frequencies spanning from $\omega$ = 850 cm$^{-1}$ to $\omega$ = 925 cm$^{-1}$. For experiments at T = 20 K, the power of the illumination is carefully maintained at 1 mW. For data collected at T = 8 K, the power is limited to 20 ~ 300 µW to reduce heating on the sample. The light is focused on an oscillating metal-coated Si tip (Akiyama tip) which generates a locally enhanced and modulated field at the tip apex [32]. This tip-enhanced field samples a localized region on the MLG with a radius of ~ 20 nm. The near-field (NF) optical information is scattered into the far-field and detected via infrared (IR) detectors. In the meantime, the tip-enhanced field heats the sample surface, driving a distribution of local photocurrent (PC) which is subsequently collected by Au contacts. The far-field background of NF and PC is suppressed via lock-in detection of the tip-modulated optical and PC signals, respectively (**see methods**).

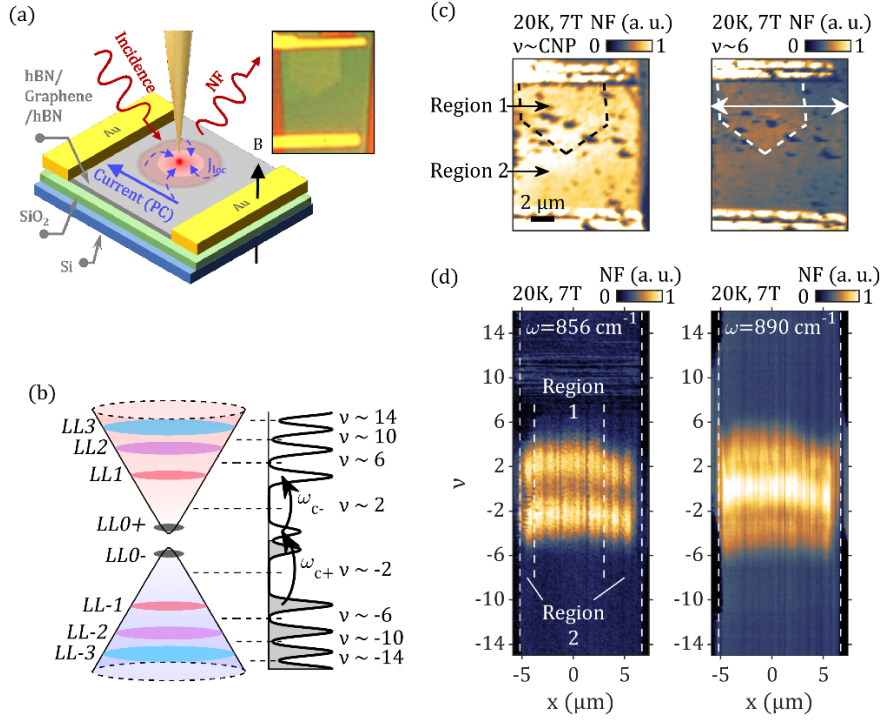

*Figure 1|**Real-space mapping of low-temperature electrodynamic and thermal dynamic properties of MLG at 7 T magnetic field.** (a) Schematics of measurement. A metal-coated tip is illuminated with the mid-IR beam. The electric field of the incident light is enhanced at the tip apex, yielding tip-scattering to the far-field as well as the local photocurrent probed via electrodes. Inset: Optical microscope image of the device layout. (b) Schematics of the density of states of graphene at low temperature and high magnetic field. At high field, the $0^{th}$ LL is split into finer sub-levels due to the lifted valley and potentially spin degeneracies. At $\nu=\pm 2, \pm 6, \pm 10…$, LLs are fully populated, and the bulk enters localized state. (c) Near-field images are acquired at $\nu \sim$CNP and $\nu \sim 6$ using $\omega=890$ cm$^{-1}$ incident beam. At CNP, MLG shows strong near-field signal due to on-resonance $0^{th} \rightarrow 1^{st}$ LL excitations. The signal is suppressed when the doping factor is increased beyond $\nu=6$ (fully occupied $1^{st}$ LL). (d) Repeated line scans are performed at a series of doping levels. The location of the line scan is indicated by the white horizontal arrow in panel (c). Line scans are acquired at $\omega=856$ cm$^{-1}$ and 890 cm$^{-1}$. Clear multi-peak structures are observed. The extended data set at more frequencies is summarized in the SI.*

## Near-field (NF) optical signal due to the $0^{th} \rightarrow 1^{st}$ Landau level transition

To precisely assign the local filling factor $\nu$ to the back gate voltages, we first examine the near-field optical response of magnetized graphene at different doping levels in **Fig. 1c** and Fig.**1d**. In **Fig. 1c**, we show NF maps of the graphene sample at T=20 K, B=7 T with an incident light frequency of $\omega=890$ cm$^{-1}$. Near CNP, we observe a strong NF response where the sample is optically 'bright' (left panel). As the sample is gated to $\nu \sim 6$, the visibility of the sample is greatly reduced (right panel). This change in optical contrast is expected as the cyclotron frequency of graphene near CNP is roughly $\omega_{c-} \sim 860$ cm$^{-1}$ and $\omega_{c+} \sim 890$ cm$^{-1}$, with a Fermi velocity $v_F \approx 1.13 \times 10^6$ m/s and a gap size $2\Delta \sim 1$ THz [33, 34]. As a result, on-resonance excitations of magnetoexciton ($0^{th} \rightarrow 1^{st}$ LL transition) at 890 cm$^{-1}$ yield strong scattering NF signal at low doping levels ($|\nu|<6$). When the $1^{st}$ LL is fully filled ($\nu \sim 6$), the NF response is obscured due to Pauli blocking. After the chemical potential enters the $2^{nd}$ LL and beyond ($\nu > 6$), graphene becomes completely dark (**Extended data in SI**).

A clearer depiction of this on-resonance to off-resonance transition is provided by the doping-dependent line scans in **Fig. 1d** (T=20 K), **Fig. S2a** (T=20K) and **Fig. S3a** in SI (T=8 K). We repeatedly scan one horizontal line across regions 1 and 2 while varying the doping level. The data collected at $\omega$=856 cm$^{-1}$≈$\omega_{c-}$, shows a clear 2-peak structure. The data collected at $\omega$=890 cm$^{-1}$≈$\omega_{c+}$, however, shows a clear 3-peak structure. The multi-peak structure and the frequency dependence agree with a split density of state of the 0$^{th}$ LL due to a lifted doubly degenerate symmetry. In the SI, we provide a phenomenological explanation using joint density of state. Importantly, this peak-like structure in the NF allows us to precisely assign our filling factor $\nu$ to the back gate voltages with a great spatial resolution. The data acquired at T=8 K (**Fig. S3 in SI**) resolves an even finer structure, confirming more lifted degeneracies at lower temperatures and high field [28, 29, 30]. We provide the full dataset at more frequency points and discuss the differences between regions 1 and 2 in **SI**. Here, we simply conclude that the doping levels of graphene in regions 1 and 2 are only slightly different, and the sample within each region is quite uniformly doped.

**Doping dependence of the nano-photocurrent (PC) signal**

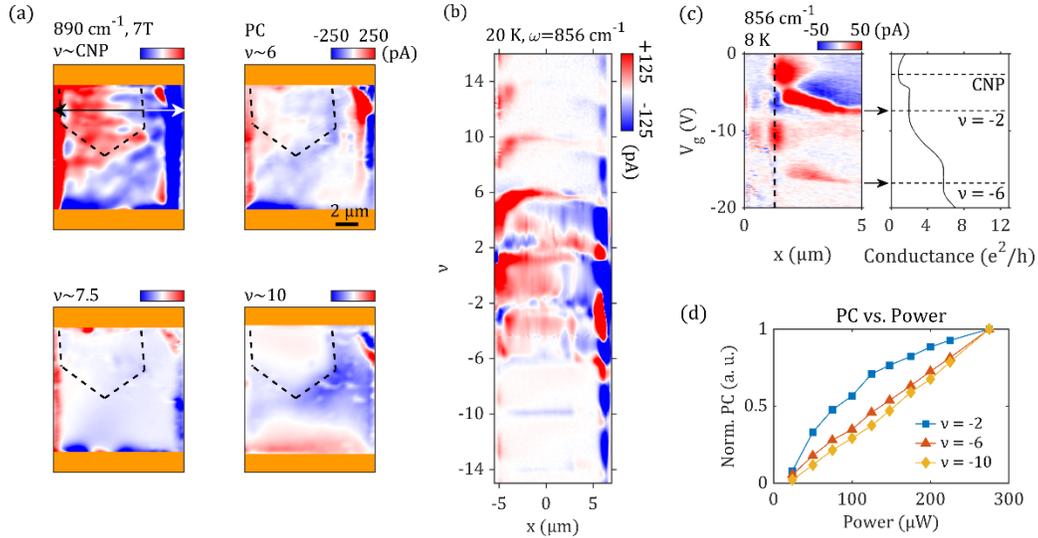

*Figure 2| **Doping-dependent photocurrent images and line scans acquired at B = 7 T.** (a) The photocurrent image near the CNP shows a strong signal accompanied by fringe-like patterns. The signal in the bulk is suppressed when 1$^{st}$ LL is fully filled ($\nu$~6). At higher doping factors, e.g. $\nu$~7.5, the photocurrent signal mainly concentrates near edges. Carefully tuning the doping factor near full-filling points of higher LLs, e.g. $\nu$~10, the signal recovers in the interior of region 2. Region 1 has a slight difference in the carrier density, resulting in different photocurrent behaviors. We note that all pseudo-color images in panel (a) have the same scale (-250 pA to 250 pA). (b) We demonstrate the doping-dependent line scans at T=20 K and $\omega$=856 cm$^{-1}$. The line scan is taken simultaneously with the optical signal in Fig. 1d. The location of the line scan is indicated by the horizontal arrow in panel (a). (c) We zoom in to the low doping range at T=8 K and compare the photocurrent line scan with a two-point conductance measurement on the same sample. We note that unlike panel (c), this line scan is taken entirely in region 2. The bulk signal is positioned exactly near the full filling points of LLs ($\nu$~-2, -6). (d) The bulk signals at $\nu$=-2, -6, and -10 are extracted at a list of power from 20 $\mu$W to 300 $\mu$W.*

The photocurrent channel provides deeper insights into the thermodynamic properties compared to the optical channels. Like the NF signal, the PC signal (**Fig. 2a**) manifests a strong response across the entire sample region near CNP. We note that an obvious fringe pattern accompanies the strong signal. The fringe pattern is likely induced by the propagating surface magnetoexciton polariton [35]. As the doping level increases from $\nu=0$ to $\nu=6$, the PC signal and the fringe pattern in the interior of the sample gradually disappear, while signals near Au edges and free edges remain. At even higher doping levels ($\nu\sim7.5$, **Fig. 2a**), the PC in the interior of the sample remains mostly zero. Interestingly, if we tune the doping factor of region 2 carefully to near $\nu\sim10$ (**Fig. 2a**), the PC signal in the entire region 2 recovers. In contrast, despite only a slight difference in doping between regions 1 and 2 ($\delta n\sim1e11$ cm$^{-2}$), no PC signal is observed in region 1. The stark contrast between the two regions highlights the narrow doping range in which the strong bulk PC signal can be found. Notably, the fringe pattern featured at $\nu<6$ is completely absent at $\nu=10$, indicating different mechanisms of generating the PC signal with and without the LL0 → LL±1 magnetoexciton excitations [35, 36, 37, 38, 39].

We acquired the doping-dependent line scans of the PC signal (**Fig. 2b**) simultaneously with the NF signal (**Fig. 1d**). Interestingly, the line scan shows a ladder-like pattern. As mentioned previously, the LL transition induced strong signal and fringe patterns populate strictly within $-6<\nu<6$. At most doping levels above the ±1$^{st}$ LL, PC signal is 'short-ranged' and concentrates at two edges of the line scan (vertical 'rails' of the 'ladder'). The edge signal becomes more prominent at half-doped LLs ($\nu\sim4n$ (n∈Z)) than fully populated LLs. However, within a narrow range of full fillings ($\nu=-2+4n$ (n∈Z)), the PC signal becomes 'long-ranged' and reoccupies the entire sample area, forming the 'rungs' of the 'ladder'. In **Fig. 2c**, we compare the PC line scan acquired at T=8 K side by side with the two-point conductance. Due to the dominance of $\sigma_{xy}$ in the two-point conductance [40], we observe clear plateau behaviors. The 'rungs' align well with the plateau and populate almost exactly near full-filling points ($\nu=-2, -6$, etc.). The clear bulk-edge correspondence, overlooked in previous far-field photocurrent experiments [41, 42, 43], can be understood within the context of the localized and extended state of a clean quantum Hall system, as we will discuss later.

To understand the role of tip-enhanced laser heating, we performed a power dependence study of the PC signal. The result is shown in **Fig. 2d**. We extracted the bulk signal near full-filling points $\nu=-2, -6$ and $-10$. The signals are normalized to the highest fluence. We find the $\nu=-2$ signal grows slower than linear dependence. In contrast, $\nu=-6$ and $-10$ signals are linear. The slower-than-linear behavior of the $\nu=-2$ signal is likely due to the strong absorption when the laser frequency is in resonance with the LL transition. The absorption leads to an increased temperature of the entire sample, rendering a deviation from the common $PC \propto |E|^2$ relation. The strict linear behaviors of signals at higher doping levels ensure that excessive heating of the entire sample is negligible in generating and measuring PC signals.

We note that due to single-electron charging effect, AFM tips can induce a current signal without photoexcitation [13, 44]. In the SI, we compared the photocurrent signal with and without illumination side-by-side. Without illumination, we observed a repeatable but weak demodulated signal. In most cases, the signal, if not zero, is much weaker than the photo-induced signal. We consider that the tip-induced signal does not influence the current discussion. The interplay between the tip-induced signal and photo-induced signal produced by weak light (<10 µW) is potentially an interesting topic for future studies.

**Bulk and Edge PC signal beyond the photo-Nernst effect**

We now focus on understanding the complete life cycle of doping-dependent photocurrent measurements in the quantum Hall regime. Previous photocurrent measurements in graphene are rather successfully explained by the hot carrier diffusion and the resulting thermoelectric effect [45]. In this scenario, carriers in graphene are heated locally by a focused laser beam or tip-enhanced field, leading to an elevated temperature $\delta T$. The hot carrier transfers heat to the lattice at a cooling rate $\gamma$ and impacts a sample area measured by the cooling length

$$l_{cool} = \sqrt{\kappa/\gamma n C_{el}} \quad (2)$$

where $\kappa$, $n$ and $C_{el}$ are the sheet thermal conductivity, carrier density and electron specific heat, respectively [45, 46]. Subsequently, local photocurrent $J_{local}$ is generated ($J_{local} = \overleftrightarrow{\alpha}\delta T$), where $\overleftrightarrow{\alpha} = \overleftrightarrow{\sigma}\overleftrightarrow{S}$ is the thermoelectric conductivity tensor that is the product of conductivity tensor $\overleftrightarrow{\sigma}$ and Seebeck and Nernst coefficients (diagonal and off-diagonal terms of $\overleftrightarrow{S}$). The local photocurrent drives the circuit, which consists of the sample and an external transimpedance amplifier, giving rise to the signal measured in the experiment. This current collecting process is summarized using the Shockly-Ramo (SR) formalism [45, 47, 48, 49].

Following the hot carrier diffusion picture, the first step is to understand the heat generation due to photon absorption. In previous magnetic-field-free experiments [46, 48, 50], the absorption primarily arises from the finite optical conductivity contributed by free electron response (plasmon absorption) and inter-band transition. Previous far-field experiments under finite magnetic field [41, 42, 43] use visible light to achieve ~micrometer spatial resolution. Due to the narrow spacing between high index LLs, the high photon energy of visible light can easily pump inter-LL transitions between LLs of high indices, contributing to a finite absorption. In our mid-IR experiment at B=7 T, the photon energy is in resonance only with the transition between the 0[th] to the 1[st] LLs. Therefore, within $|\nu|<6$, the LL transition leads to a pronounced optical absorption. At $|\nu|>6$, the frequency of the dominant transition (n → n+1 LL) is lower than the incident photon energy. The optical conductivity due to this transition is [51],

$$\sigma_{xx} = \frac{e^3 v_F^2 B}{\omega_c \pi c i} \frac{\omega + i\Gamma}{\omega_c^2 - (\omega + i\Gamma)^2}$$

This conductivity describes a branch of magneto-plasmon. Despite no on-resonance optical transition, the magneto-plasmon response at the higher frequency end of the resonance dictates the absorption, albeit weaker. In the limit of low magnetic field and even higher doping, the dispersion of magneto-plasmon reduces to $\omega(\boldsymbol{q}) = \sqrt{\omega_c^2 + \omega_p^2(\boldsymbol{q})}$. Here, $\omega_p(\boldsymbol{q})$ is the field-free plasmon dispersion. Therefore, our AFM tip, capable of coupling to modes with a range of momentum, induces a finite absorption through magneto-plasmon response.

We now analyze the "ladder" structure in our nanoscale PC measurements in comparison to previous microscale photocurrent measurements [41, 42, 43]. To further illustrate the essential real-space features, we took two line-cuts in the $\omega$=890 cm[-1] data set (**Fig. 3a**) at the edge and bulk of the sample. Here, we limit our discussion to high doping levels $|\nu|>6$, where we avoid enhanced PC signal and sample temperature due to contributions from the on-resonance magnetoexciton excitations. As mentioned above, we see that near full fillings ($\nu$=-2+4n (n∈Z)), the bulk PC signal is non-zero, while around half-fillings ($\nu$=4n (n∈Z)) the PC signal at the edge is non-zero. In the **SI**, we show more data following strictly the same trend at higher LLs. This bulk/edge correspondence is counterintuitive and has not been previously addressed: both Seebeck and Nernst coefficient should be zero near full fillings ($\sigma_{xy}$ plateaus) where no free charge remains in the sample (**Fig.3b**) [7, 8]. Therefore, the presence of bulk PC signal at full fillings is nontrivial. In addition, we found that the edge PC is mainly single-signed at $|\nu|>6$ (e.g., in **Fig. 2b** and **Fig. 3a**, positive on left edge and negative on the right edge), similar to the observation in the previous far-field experiment [41].

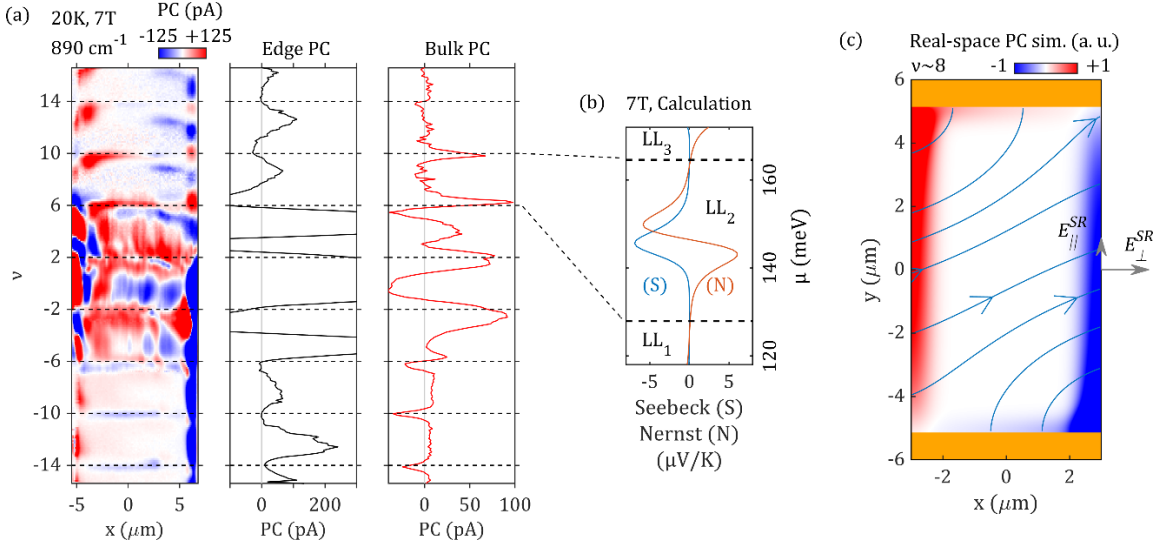

*Figure 3|**Nano-photocurrent line-cuts on the edge and in the bulk.** (a) Nano-photocurrent shows different behaviors on the free edge and in the bulk of the sample. Two doping dependent curves representing the edge (x~-5 µm) and the bulk (x ~ -1 µm) are extracted from the doping dependent line scan. (b) We plot chemical potential (µ) dependent thermopower coefficients. Specifically, we demarcate the 1$^{st}$, 2$^{nd}$ and the 3$^{rd}$ LL using dashed lines. (c) Following the Shockly-Ramo formalism, the auxiliary field (blue arrow curves) $E^{SR}$ is computed on a rectangular MLG sample doped near $\nu$~8. The photocurrent map is subsequently simulated.*

**Edge PC signal at half Landau level fillings**

To understand the real-space pattern of PC at half fillings, we perform simulations using the SR formalism [45, 47, 48, 49] (**Details in SI**). When LLs are half-filled ($\nu$=4, 8 or 12), the chemical potential is set at extended states, rendering the interior of the sample metallic. Consequently, the auxiliary field $\boldsymbol{E}_{SR} = -\nabla \phi^{SR}$ established over the entire sample remains a valid assumption, as constructed in the SR formalism for characterizing the hot carrier diffusion under a local temperature gradient [47]. Here, the quantum Hall edge states are ignored due to the metallic bulk. Under illumination, the tip generates a nanoscale temperature profile $\delta T \propto \sigma_1(\omega, q) |E_{inc}|^2$. The signal can be computed as

$$I_{PC} \propto \int \nabla \delta T \cdot \vec{\alpha}^T \cdot \nabla \phi^{SR} dr.$$

We focus on the free edge of an ideally uniform sample. By converting the bulk integration to boundary integration and applying boundary condition of $j_n^{SR}$=0 we have,

$$I_{PC} \propto \int_{\partial S} \alpha_{xx} \delta T d\boldsymbol{n} \cdot \nabla \phi_{SR} - \int_{\partial S} \alpha_{xy} \delta T (d\boldsymbol{n} \times \nabla \phi_{SR})_z$$

$$\propto \pm(\alpha_{xx}\sigma_{xy} - \alpha_{xy}\sigma_{xx})\int \delta T dl \propto \pm S_{xy}\int \delta T dl$$

Here, the result reduces to the Nernst coefficient $S_{xy}$. The $\pm$ sign depends on whether the heating happens on the left-side or the right-side free edge. We note that $S_{xy}$ switches sign when the chemical potential sweeps across one LL (**Fig. 3b**) [7, 8]. This contradicts our observation that the doping-dependent signal on the free edge has a clear single-sign bias (**Fig. 3a**). In the previous far-field PC experiment [41], similar behavior was observed. However, due to a lack of fine resolution (~micrometer), the contradiction was attributed to an unknown background. With nano-photocurrent technique, we observed this phenomenon persists even though we have a much finer resolution (~20 nm). Therefore, despite the fact that hot carriers might trigger the current flow in the quantum Hall extended state, the edge photocurrent cannot be simply described by carrier thermal diffusion in a metal with Hall conductivity. The single-signed bias of edge photocurrent likely indicates its origin from the chiral edge states of the quantum Hall regime [42].

For a clean sample doped at half-fillings, the photocurrent is evident on the free edge of the sample (e.g. data in **Fig. 2a**). Here the decay length of the signal near the edge is gauged by the cooling length $l_{cool}$. In **Fig. 3a**, we found $l_{cool} \approx 200$ nm near $\nu$~8. In **Fig. 3c**, we reproduce the photocurrent signal at the free edges of graphene at partially filled LLs under SR formalism.

**Bulk PC signal at integer Landau level fillings**

Near the full filling points of LLs $\nu$=-2+4n (n∈Z), electronic states tend to be localized around the disorder, preventing current flow in the interior of sample. This results in a vanishingly small, if not zero, longitudinal conductivity, whereas Hall conductivity is quantized into plateaus (**Fig. 4a**). The thermal fluctuation at 8 K is also tiny compared to LL gaps (e.g., at $\nu = 2$, $k_B T/\hbar\omega_c$<0.01), due to the exceptionally high Fermi velocity of graphene. As a result, carrier diffusion is strongly suppressed unless the diffusion happens on the edge channel. This requires tip-induced heating to reach the Au electrodes or free edges of graphene to yield a current flow. Therefore, the existence of a strong PC signal in the entire bulk of a 10-micrometer sample indicates a strongly enhanced cooling length compared to that of the extended state. The long cooling length $l_{cool} = \sqrt{\kappa_{xx}/n\gamma C_{el}}$, which is hidden in far-field PC experiments, anticipates either a strong anomaly in the thermal conductivity or a thermally isolated electron and lattice dynamics in the localized state.

In metals, thermal and electrical conductivities are typically related by the WF law, $\kappa_{xx}/\sigma_{xx} = LT$, where the two conductivities are proportional with a constant Lorentz number $L$. This approximation expects a vanishing thermal conductivity in the quantum Hall localized state. Despite being a 2D semimetal, theories [52, 21] predict that the WF law is usually not violated in graphene at the zero-temperature limit. In other words, like the electrical conductivity $\sigma_{xx}$, the thermal conductivity $\kappa_{xx}$ should decrease much faster than the carrier density $n$. If we temporarily neglect the doping dependence of $\gamma$, the $\sqrt{\kappa_{xx}/n}$ leads to a strong trend of vanishing cooling length at full LL fillings, which contradicts our observation. Previous experiments on semiconductor mesa of GaAs–AlGaAs heterostructure revealed that WF law is indeed violated in the localized state of the $\nu$~2 quantum Hall plateau [10]. The electrical conductivity $\sigma_{xx}$ is sharply suppressed when entering the localized state, while the thermal conductivity $\kappa_{xx}$ decays only in a power-law trend with magnetic field.

In the case of graphene under high magnetic field, theories indeed predicted a much stronger violation of WF law [20, 21, 22] than that of semiconductors at finite temperatures. The electrical and thermal conductivities are expected to exhibit 'anti-phase' quantum oscillation when sweeping across LLs. Contrary to the WF law, the thermal conductivity is predicted to be relatively lower in the extended state and much enhanced in the localized state. This enhancement leads to an extended cooling length, allowing the tip-induced heat to propagate from the sample's interior to its boundary, potentially supporting the strong violation of WF law.

In addition to the thermal conductivity, Bistritzer and MacDonald [53] predict that the electron cooling rate to the lattice ($\gamma$) in graphene varies with carrier density. A full description of the cooling rate in the quantum Hall regime is not yet available, however, to make a plausible statement, we can estimate the cooling rate in the quantum hall graphene by comparing it to the field-free case with a similar carrier density. For half-filled LLs at 7T, the carrier density is comparable to a field-free graphene with a chemical potential of $\mu \approx 70$ meV. The electron cooling time in the extended state is estimated to be in the order of ~2 ns. In the case of full-filling LLs with a lattice temperature $T_L=8$ K, the cooling time is dramatically increased to the order of 700 ns to 11 $\mu$s. Therefore, this increase in the cooling time contributes to the enhancement of the cooling length by at least an order of magnitude. The ultra-long cooling time scale in the localized state also supported thermally isolated electron and lattice dynamics. The isolation indicates the thermal conduction through graphene lattice and substrate (hBN) plays a negligible role in the localized quantum Hall state. More importantly, this isolation explains the unexpectedly strong signal at full-fillings observed only with tip-based optical excitation, with which graphene electrons can be heated due to magneto-plasmon absorption. In far-field photocurrent measurements, momentum mismatch suppressed the absorption in the bulk. In DC measurements, graphene electrons cannot be easily heated via thermal transport through the lattice vibrations. This isolation also highlights that, despite the electronic diffusion being sharply suppressed in the localized state, thermal conduction is not forbidden.

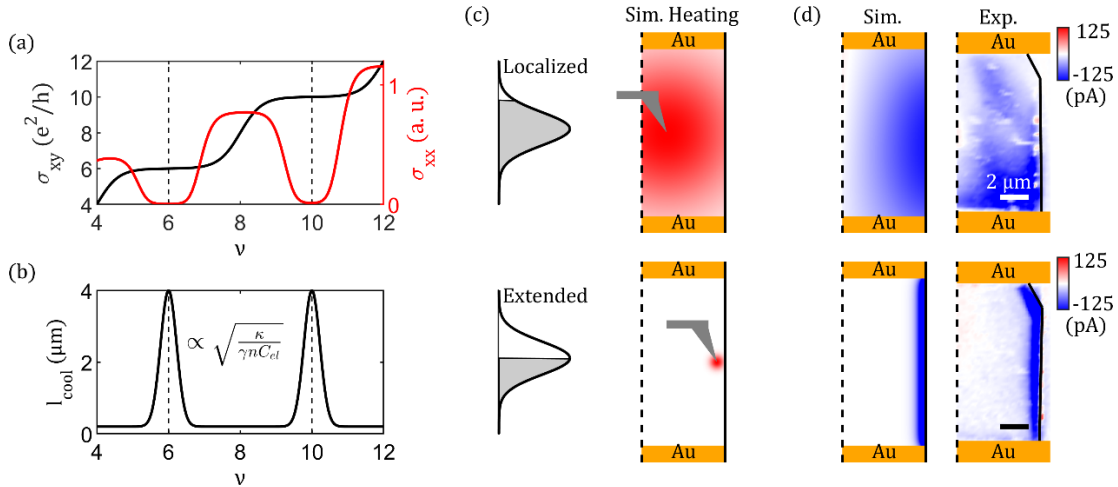

*Figure 4| **Photocurrent at full filling points of LLs and the violation of Wiedemann-Franz law.** (a) At low temperature, Hall conductivity of graphene shows clear plateaus in the localized states. Longitudinal conductivity vanishes in localized states. (b) The cooling length $l_{cool}$ exhibits an anomalous enhancement in the localized state compared to the extended state. (c) Nanoscale heating induced by the tip-enhanced electric field spread onto a graphene half plane characterized by the cooling length. The heating profile is schematically illustrated in the localized and the extended state. The dashed line indicates the continued sample region, while the solid line marks the physical sample*

*edge. (d) When the tip-induced heating reaches the free edges and Au edges of the sample, carrier diffusion on the chiral edge channel contributes to an observable current signal. Due to the strongly enhanced cooling length in the localized state, the signal spreads on the entire interior of the sample.*

**In Fig. 4b**, we estimate the cooling length dependence on the doping factor. In the extended regime, the cooling length is short (~200 nm) and therefore the calculated PC signal can be compared to the data (bottom panel of **Fig. 4c** and **Fig. 4d**). When entering the localized state, the cooling length dramatically increases, which is comparable to or larger than the length of the sample (upper panel of **Fig. 4c** and **Fig. 4d**). As a result, this nonlocal heating reaches contacts or free edges where chiral hot carrier diffusion can take place and ultimately contributes to the collected photocurrent signal.

All the essential features discussed above at the bulk and edge of graphene have been confirmed in other samples and at THz frequencies. In the SI, we summarized doping dependence of PC using $\omega$=0.8 THz illumination. THz near-field PC reveals bulk signal at doping factors $\nu$=-1, 0 and 1 together with a similar edge-bulk correspondence ("rungs" and "rails" structures). One difference with THz illumination is that, at high magnetic field and low doping, the cyclotron frequency is much higher than the THz photon energy. Therefore, bulk magneto-plasmon is not likely responsible for thermal absorption as in the IR case. However, strong photocurrent signals are still observed at the edge of the sample in a broad range of doping levels near the CNP up to B=7 T. In this scenario, the magneto edge plasmon (MEP) is likely responsible for the absorption. MEP is the un-gapped acoustic plasmonic mode running along the edge of graphene [54, 55, 56], which can also be coupled by the near-field tip. This is a potentially interesting topic for future studies.

We also note that the PC signal near full-fillings shares many similarities with the strong signal near CNP in field-free measurements [45, 48, 57]. In the latter case, the nano-photocurrent is exceptionally pronounced and, like the former case, exhibits an extended cooling length. Near the CNP, the strong Seebeck effect at zero magnetic field is due to two main factors: a strong imbalance of the DOS above and below the chemical potential [58] and the low electrical conductivity [58]. The cooling length is extended compared to high gating levels due to the reduced carrier density and inefficient thermal exchange between the electron and the lattice [59, 45]. Despite many similarities between the two cases, the localized nature of the cyclotron orbits, transverse electrical and thermoelectrical responses, and the bulk/edge correspondence of the quantum hall state distinguish the bulk signal observed near full LL fillings from the field-free signal.

**Conclusion**

In conclusion, we have, for the first time, studied simultaneously the optical and photothermal properties of graphene with high spatial resolution in the deep quantum Hall regime down to 8 K and 7 T. The power-dependence studies ensure that laser-induced heating minimally impacts higher LLs, confirming that the observed PC signals are intrinsic to quantum Hall physics. This experimental capability enables Landau level nanoscopy and nano-spectroscopy in magnetized electron gases, offering opportunities to investigate magneto-optical and thermodynamic properties at the nanoscale. In a clean graphene device, we observe strong PC signals particularly near full fillings ($\nu = -2 + 4n$), where the PC signal extends across the entire sample bulk, forming a "ladder-like" structure with a clear bulk-edge correspondence in gate-dependent measurements. This long-range PC signal suggests an enhanced cooling length in the bulk, challenging the WF law, which predicts a vanishing thermal conductivity in the localized state. Additionally, at half-fillings ($v \approx 4n$), the PC signal again reveals distinct behaviors at the edge and bulk, with pronounced non-Nernst-like signals at the edge. Near CNP, the identification of valley and spin splitting in the zeroth LL further highlights the role of degeneracy lifting and magnetoplasmons in shaping nanoscale thermal transport. Future studies can leverage SNOM and photothermal nanoscopy to explore emerging quantum phases with nontrivial charge density variations and edge channels. For example, optical and thermal effects associated with electrons in flat bands [60], fractionalized quasiparticles, unconventional superconductivity, as well as light-matter interactions in magnetized 2D cavities [61, 62] can be readily studied.

**Method:**

The tip oscillates at ~50 kHz, enabling tapping mode AFM and encoding the NF and PC signal at integer harmonics of the tapping frequency. In a tip-based NF experiment, far-field effects due to the finite size ($\lambda/2$~5 um) of the focused beam, causing unwanted unlocalized signal, are inevitable. The higher harmonics of NF and PC signal efficiently suppress the far-field component [63]. In mid-IR experiments, the 4$^{th}$ harmonics of NF and 2$^{nd}$ harmonics of PC are used to achieve the localized signal.

**Acknowledgement**

We are grateful to Michael M. Fogler (University of California San Diego), Justin C. W. Song (Nanyang Technological University), Philip B. Allen (Stony Brook University) and Keji Lai (University of Texas at Austin) for helpful discussion.

**Funding:** R. J., Q. L., and M. K. L. acknowledge support from the U.S. Department of Energy, Office of Basic Energy Sciences, Division of Material Science and Engineering, under contract No. DE-SC0012704 for the support of chiral transport. Research on m-SNOM scanning probe platform is supported as part of Programmable Quantum Materials, an Energy Frontier Research Center funded by the U.S. Department of Energy (DOE), Office of Science, Basic Energy Sciences (BES), under award DE-SC0019443. M.K.L. acknowledges Gordon and Betty Moore Foundation DOI: 10.37807/gbmf12258 for supporting the development of polaritonic materials. L.W., M. K. Liu and D. N. Basov acknowledge support from the U.S. Department of Energy, Office of Science, National Quantum Information Science Research Centers, Co-design Center for Quantum Advantage (C2QA) under contract number DE-SC0012704 for support of the data analysis. The US Government retains a non-exclusive, paid-up, irrevocable, world-wide license to publish or reproduce the published form of this paper, or allow others to do so, for US Government purposes. DOE will provide public access (http://energy.gov/downloads/doe-public-access-plan). Distributed under a Creative Commons Attribution NonCommercial License 4.0 (CC BY-NC).

**Author contributions:**

R.J. and M.K.L. conceived the project and designed the experiments. G.L.C., X.D., D.N.B., Q.L. and M.K.L. supervised the project. B.Z., J.S., S.C. and X.D. prepared the devices. R.J. performed the experiments with the support of S.C., W.Z., Z.Z., H.W., M.D., Y.D. and Z.D. R.J. and M.K.L. analyzed experimental data with the input from B.Z., J.S., S.C., L.W., B.C. and Q.L.. R.J. and

M.K.L. developed numerical simulation and theoretical description. R.J. and M.K.L. cowrote the manuscript with input from all co-authors.

**Competing interests:** M.D., X.C., and M.K.L. have filed a patent regarding magneto scanning near-field optical microscopy (publication number: WO/2023/049225; publication date: 30 March 2023; international application no.: PCT/US2022/044311; international filing date: 22 September 2022; organizations issuing patent: The Research Foundation For The State University Of New York, 35 State Street, Albany, New York 12207, USA; Yale University, Two Whitney Ave, New Haven, Connecticut 06511, USA). The other authors declare that they have no competing interests.

**Data and materials availability:** The data that support the findings of this study are available from the corresponding author upon reasonable request.

# Bibliography

[1]  R. B. Laughlin, "Quantized Hall conductivity in two dimensions," *Physical Review B,* vol. 23, p. 5632, 1981.

[2]  B. I. Halperin, "Quantized Hall conductance, current-carrying edge states, and the existence of extended states in a two-dimensional disordered potential," *Physical review B,* vol. 25, p. 2185, 1982.

[3]  D. C. Tsui, H. L. Störmer and A. C. Gossard, "Zero-resistance state of two-dimensional electrons in a quantizing magnetic field," *Physical Review B,* vol. 25, p. 1405, 1982.

[4]  K. S. Novoselov, A. K. Geim, S. V. Morozov, D.-e. Jiang, Y. Zhang, S. V. Dubonos, I. V. Grigorieva and A. A. Firsov, "Electric field effect in atomically thin carbon films," *science,* vol. 306, p. 666–669, 2004.

[5]  Y. Zhang, Y.-W. Tan, H. L. Stormer and P. Kim, "Experimental observation of the quantum Hall effect and Berry's phase in graphene," *nature,* vol. 438, p. 201–204, 2005.

[6]  A. K. Geim and K. S. Novoselov, "The rise of graphene," *Nature materials,* vol. 6, p. 183–191, 2007.

[7]  J. G. Checkelsky and N. P. Ong, "Thermopower and Nernst effect in graphene in a magnetic field," *Physical Review B—Condensed Matter and Materials Physics,* vol. 80, p. 081413, 2009.

[8]  Y. M. Zuev, W. Chang and P. Kim, "Thermoelectric and magnetothermoelectric transport measurements of graphene," *Physical review letters,* vol. 102, p. 096807, 2009.

[9] R. Fletcher, J. C. Maan, K. Ploog and G. Weimann, "Thermoelectric properties of GaAs-Ga$_{1-x}$Al$_x$As heterojunctions at high magnetic fields," *Physical Review B,* vol. 33, p. 7122, 1986.

[10] R. A. Melcer, A. Gil, A. K. Paul, P. Tiwari, V. Umansky, M. Heiblum, Y. Oreg, A. Stern and E. Berg, "Heat conductance of the quantum Hall bulk," *Nature,* vol. 625, p. 489–493, 2024.

[11] J. Waissman, L. E. Anderson, A. V. Talanov, Z. Yan, Y. J. Shin, D. H. Najafabadi, M. Rezaee, X. Feng, D. G. Nocera, T. Taniguchi and others, "Electronic thermal transport measurement in low-dimensional materials with graphene non-local noise thermometry," *Nature Nanotechnology,* vol. 17, p. 166–173, 2022.

[12] J. P. Pekola and B. Karimi, "Colloquium: Quantum heat transport in condensed matter systems," *Reviews of Modern Physics,* vol. 93, p. 041001, 2021.

[13] D. Halbertal, J. Cuppens, M. B. Shalom, L. Embon, N. Shadmi, Y. Anahory, H. R. Naren, J. Sarkar, A. Uri, Y. Ronen and others, "Nanoscale thermal imaging of dissipation in quantum systems," *Nature,* vol. 539, p. 407–410, 2016.

[14] F. Menges, P. Mensch, H. Schmid, H. Riel, A. Stemmer and B. Gotsmann, "Temperature mapping of operating nanoscale devices by scanning probe thermometry," *Nature communications,* vol. 7, p. 10874, 2016.

[15] M. Mecklenburg, W. A. Hubbard, E. R. White, R. Dhall, S. B. Cronin, S. Aloni and B. C. Regan, "Nanoscale temperature mapping in operating microelectronic devices," *Science,* vol. 347, p. 629–632, 2015.

[16] Y. Yue and X. Wang, "Nanoscale thermal probing," *Nano reviews,* vol. 3, p. 11586, 2012.

[17] S. Sadat, A. Tan, Y. J. Chua and P. Reddy, "Nanoscale thermometry using point contact thermocouples," *Nano letters,* vol. 10, p. 2613–2617, 2010.

[18] Y. Zhang, W. Zhu, F. Hui, M. Lanza, T. Borca-Tasciuc and M. Muñoz Rojo, "A review on principles and applications of scanning thermal microscopy (SThM)," *Advanced functional materials,* vol. 30, p. 1900892, 2020.

[19] F. Giazotto, T. T. Heikkilä, A. Luukanen, A. M. Savin and J. P. Pekola, "Opportunities for mesoscopics in thermometry and refrigeration: Physics and applications," *Reviews of Modern Physics,* vol. 78, p. 217–274, 2006.

[20] V. P. Gusynin and S. G. Sharapov, "Magnetic oscillations in planar systems with the Dirac-like spectrum of quasiparticle excitations. II. Transport properties," *Physical Review B—Condensed Matter and Materials Physics,* vol. 71, p. 125124, 2005.

[21] B. Dóra and P. Thalmeier, "Magnetotransport and thermoelectricity in Landau-quantized disordered graphene," *Physical Review B—Condensed Matter and Materials Physics,* vol. 76, p. 035402, 2007.

[22] W. Duan, J.-F. Liu, C. Zhang and Z. Ma, "Thermoelectric and thermal transport properties of graphene under strong magnetic field," *Physica E: Low-dimensional Systems and Nanostructures,* vol. 104, p. 173–176, 2018.

[23] K. Ulrich and P. Esquinazi, "Magnetothermal transport of oriented graphite at low temperatures," *Journal of low temperature physics,* vol. 137, p. 217–231, 2004.

[24] Z. Pan, G. Lu, X. Li, J. R. McBride, R. Juneja, M. Long, L. Lindsay, J. D. Caldwell and D. Li, "Remarkable heat conduction mediated by non-equilibrium phonon polaritons," *Nature,* vol. 623, p. 307–312, 2023.

[25] J. G. Checkelsky, L. Li and N. P. Ong, "Zero-energy state in graphene in a high magnetic field," *Physical review letters,* vol. 100, p. 206801, 2008.

[26] E. A. Henriksen, P. Cadden-Zimansky, Z. Jiang, Z. Q. Li, L.-C. Tung, M. E. Schwartz, M. Takita, Y.-J. Wang, P. Kim and H. L. Stormer, "Interaction-induced shift of the cyclotron resonance of graphene using infrared spectroscopy," *Physical review letters,* vol. 104, p. 067404, 2010.

[27] Y. Zhang, Z. Jiang, J. P. Small, M. S. Purewal, Y.-W. Tan, M. Fazlollahi, J. D. Chudow, <. f. J. A. Jaszczak, H. L. Stormer and P. Kim, "Landau-level splitting in graphene in high magnetic fields," *Physical review letters,* vol. 96, p. 136806, 2006.

[28] J. Alicea and M. P. A. Fisher, "Graphene integer quantum Hall effect in the ferromagnetic and paramagnetic regimes," *Physical Review B—Condensed Matter and Materials Physics,* vol. 74, p. 075422, 2006.

[29] A. F. Young, C. R. Dean, L. Wang, H. Ren, P. Cadden-Zimansky, K. Watanabe, T. Taniguchi, J. Hone, K. L. Shepard and P. Kim, "Spin and valley quantum Hall ferromagnetism in graphene," *Nature Physics,* vol. 8, p. 550–556, 2012.

[30] X. Du, I. Skachko, F. Duerr, A. Luican and E. Y. Andrei, "Fractional quantum Hall effect and insulating phase of Dirac electrons in graphene," *Nature,* vol. 462, p. 192–195, 2009.

[31] T. Han, J. Yang, Q. Zhang, L. Wang, K. Watanabe, T. Taniguchi, P. L. McEuen and L. Ju, "Accurate measurement of the gap of graphene/h-BN moiré superlattice through photocurrent spectroscopy," *Physical Review Letters,* vol. 126, p. 146402, 2021.

[32] B. Knoll and F. Keilmann, "Enhanced dielectric contrast in scattering-type scanning near-field optical microscopy," *Optics communications,* vol. 182, p. 321–328, 2000.

[33] Y. J. Song, A. F. Otte, Y. Kuk, Y. Hu, D. B. Torrance, P. N. First, W. A. de Heer, H. Min, S. Adam, M. D. Stiles and others, "High-resolution tunnelling spectroscopy of a graphene quartet," *Nature,* vol. 467, p. 185–189, 2010.

[34] N. R. Finney, M. Yankowitz, L. Muraleetharan, K. Watanabe, T. Taniguchi, C. R. Dean and J. Hone, "Tunable crystal symmetry in graphene–boron nitride heterostructures with coexisting moiré superlattices," *Nature nanotechnology,* vol. 14, p. 1029–1034, 2019.

[35] M. Dapolito, M. Tsuneto, W. Zheng, L. Wehmeier, S. Xu, X. Chen, J. Sun, Z. Du, Y. Shao, R. Jing and others, "Infrared nano-imaging of Dirac magnetoexcitons in graphene," *Nature Nanotechnology,* vol. 18, p. 1409–1415, 2023.

[36] L. Wehmeier, S. Xu, R. A. Mayer, B. Vermilyea, M. Tsuneto, M. Dapolito, R. Pu, Z. Du, X. Chen, W. Zheng and others, "Nano-imaging of Landau-phonon polaritons in Dirac heterostructures," *arXiv preprint arXiv:2312.14093,* 2023.

[37] V. P. Gusynin, S. G. Sharapov and J. P. Carbotte, "Magneto-optical conductivity in graphene," *Journal of Physics: Condensed Matter,* vol. 19, p. 026222, 2006.

[38] I. O. Nedoliuk, S. Hu, A. K. Geim and A. B. Kuzmenko, "Colossal infrared and terahertz magneto-optical activity in a two-dimensional Dirac material," *Nature nanotechnology,* vol. 14, p. 756–761, 2019.

[39] A. Rikhter, D. N. Basov and M. M. Fogler, "Modeling of plasmonic and polaritonic effects in photocurrent nanoscopy," *Journal of Applied Physics,* vol. 135, 2024.

[40] J. R. Williams, D. A. Abanin, L. DiCarlo, L. S. Levitov and C. M. Marcus, "Quantum Hall conductance of two-terminal graphene devices," *Physical Review B—Condensed Matter and Materials Physics,* vol. 80, p. 045408, 2009.

[41] H. Cao, G. Aivazian, Z. Fei, J. Ross, D. H. Cobden and X. Xu, "Photo-Nernst current in graphene," *Nature Physics,* vol. 12, p. 236–239, 2016.

[42] B. Cao, T. Grass, O. Gazzano, K. A. Patel, J. Hu, M. Muller, T. Huber-Loyola, L. Anzi, K. Watanabe, T. Taniguchi and others, "Chiral transport of hot carriers in graphene in the quantum Hall regime," *ACS nano,* vol. 16, p. 18200–18209, 2022.

[43] G. Nazin, Y. Zhang, L. Zhang, E. Sutter and P. Sutter, "Visualization of charge transport through Landau levels in graphene," *Nature Physics,* vol. 6, p. 870–874, 2010.

[44] M. T. Woodside and P. L. McEuen, "Scanned probe imaging of single-electron charge states in nanotube quantum dots," *Science,* vol. 296, p. 1098–1101, 2002.

[45] J. C. W. Song, M. S. Rudner, C. M. Marcus and L. S. Levitov, "Hot carrier transport and photocurrent response in graphene," *Nano letters,* vol. 11, p. 4688–4692, 2011.

[46] A. Woessner, P. Alonso-González, M. B. Lundeberg, Y. Gao, J. E. Barrios-Vargas, G. Navickaite, Q. Ma, D. Janner, K. Watanabe, A. W. Cummings and others, "Near-field photocurrent nanoscopy on bare and encapsulated graphene," *Nature communications,* vol. 7, p. 10783, 2016.

[47] J. C. W. Song and L. S. Levitov, "Shockley-Ramo theorem and long-range photocurrent response in gapless materials," *Physical Review B,* vol. 90, p. 075415, 2014.

[48] Q. Ma, C. H. Lui, J. C. W. Song, Y. Lin, J. F. Kong, Y. Cao, T. H. Dinh, N. L. Nair, W. Fang, K. Watanabe and others, "Giant intrinsic photoresponse in pristine graphene," *Nature nanotechnology,* vol. 14, p. 145–150, 2019.

[49] Q. Ma, R. Krishna Kumar, S.-Y. Xu, F. H. L. Koppens and J. C. W. Song, "Photocurrent as a multiphysics diagnostic of quantum materials," *Nature Reviews Physics,* vol. 5, p. 170–184, 2023.

[50] Y. Shao, R. Jing, S. H. Chae, C. Wang, Z. Sun, E. Emmanouilidou, S. Xu, D. Halbertal, B. Li, A. Rajendran and others, "Nonlinear nanoelectrodynamics of a Weyl metal," *Proceedings of the National Academy of Sciences,* vol. 118, p. e2116366118, 2021.

[51] V. P. Gusynin and S. G. Sharapov, "Transport of Dirac quasiparticles in graphene: Hall and optical conductivities," *Physical Review B—Condensed Matter and Materials Physics,* vol. 73, p. 245411, 2006.

[52] Y. M. Blanter, D. V. Livanov and M. O. Rodin, "Thermal conductivity in the quantum Hall effect regime," *Journal of Physics: Condensed Matter,* vol. 6, p. 1739, 1994.

[53] R. Bistritzer and A. H. MacDonald, "Electronic cooling in graphene," *Physical Review Letters,* vol. 102, p. 206410, 2009.

[54] I. Crassee, M. Orlita, M. Potemski, A. L. Walter, M. Ostler, T. Seyller, I. Gaponenko, J. Chen and A. B. Kuzmenko, "Intrinsic terahertz plasmons and magnetoplasmons in large scale monolayer graphene," *Nano letters,* vol. 12, p. 2470–2474, 2012.

[55] H. Yan, Z. Li, X. Li, W. Zhu, P. Avouris and F. Xia, "Infrared spectroscopy of tunable Dirac terahertz magneto-plasmons in graphene," *Nano letters,* vol. 12, p. 3766–3771, 2012.

[56] V. A. Volkov and S. A. Mikhailov, "Edge magnetoplasmons: low frequency weakly damped excitations in inhomogeneous two-dimensional electron systems," *Sov. Phys. JETP,* vol. 67, p. 1639–1653, 1988.

[57] D. Wang, A. E. L. Allcca, T.-F. Chung, A. V. Kildishev, Y. P. Chen, A. Boltasseva and V. M. Shalaev, "Enhancing the graphene photocurrent using surface plasmons and a pn junction," *Light: Science & Applications,* vol. 9, p. 126, 2020.

[58] J. Duan, X. Wang, X. Lai, G. Li, K. Watanabe, T. Taniguchi, M. Zebarjadi and E. Y. Andrei, "High thermoelectricpower factor in graphene/hBN devices," *Proceedings of the National Academy of Sciences,* vol. 113, p. 14272–14276, 2016.

[59] A. Poux, Z. R. Wasilewski, K. J. Friedland, R. Hey, K. H. Ploog, R. Airey, P. Plochocka and D. K. Maude, "Microscopic model for the magnetic-field-driven breakdown of the dissipationless state in the integer and fractional quantum Hall effect," *Physical Review B,* vol. 94, p. 075411, 2016.

[60] S. Batlle-Porro, D. Calugaru, H. Hu, R. K. Kumar, N. C. H. Hesp, K. Watanabe, T. Taniguchi, B. A. Bernevig, P. Stepanov and F. H. L. Koppens, "Cryo-Near-Field Photovoltage Microscopy of Heavy-Fermion Twisted Symmetric Trilayer Graphene," *arXiv preprint arXiv:2402.12296,* 2024.

[61] V. Rokaj, M. Penz, M. A. Sentef, M. Ruggenthaler and A. Rubio, "Quantum electrodynamical Bloch theory with homogeneous magnetic fields," *Physical review letters,* vol. 123, p. 047202, 2019.

[62] V. Rokaj, M. Penz, M. A. Sentef, M. Ruggenthaler and A. Rubio, "Polaritonic Hofstadter butterfly and cavity control of the quantized Hall conductance," *Physical Review B,* vol. 105, p. 205424, 2022.

[63] A. J. Sternbach, J. Hinton, T. Slusar, A. S. McLeod, M. K. Liu, A. Frenzel, M. Wagner, R. Iraheta, F. Keilmann, A. Leitenstorfer and others, "Artifact free time resolved near-field spectroscopy," *Optics Express,* vol. 25, p. 28589–28611, 2017.

# Photocurrent Nanoscopy of Quantum Hall Bulk

Ran Jing[1*], Boyi Zhou[2,3], Jiacheng Sun[2], Shoujing Chen[2], Wenjun Zheng[2], Zijian Zhou[2], Heng Wang[2], Lukas Wehmeier[2,5], Bing Cheng[2], Michael Dapolito[2,4], Yinan Dong[4], Zengyi Du[2], G. L. Carr[5], Xu Du[2], D. N. Basov[4], Qiang Li[1,2], Mengkun Liu[2,5*]

[1]*Condensed Matter Physics and Materials Science Division, Brookhaven National Laboratory; Upton, NY, 11973-5000, USA.*

[2] *Department of Physics and Astronomy, Stony Brook University; Stony Brook, New York 11794, USA.*

[3]*Center for Integrated Science and Engineering, Columbia University, New York, New York 10027, USA*

[4]*Department of Physics, Columbia University; New York, New York 10027, USA.*

[5]*National Synchrotron Light Source II, Brookhaven National Laboratory; Upton, New York 11973, USA.*

*rjing@bnl.gov

*mengkun.liu@stonybrook.edu

**Section 1: Doping and frequency dependence of the nano-optical and nano-photocurrent data**

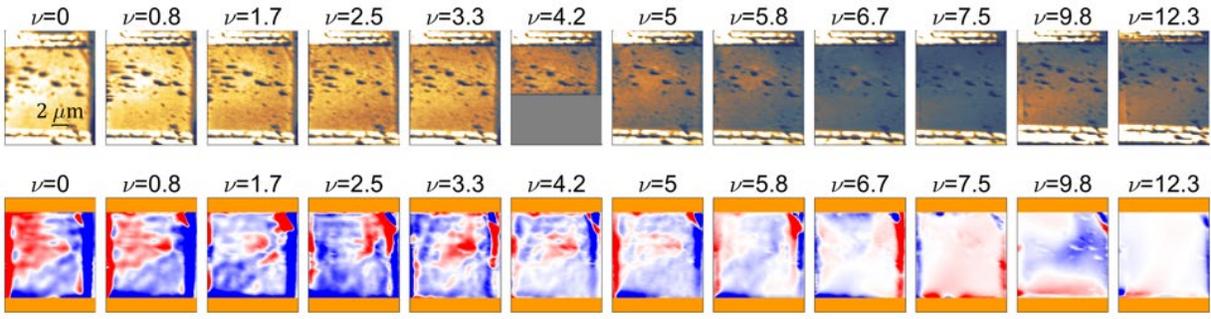

*Supplementary Figure 1| Doping-dependent nano-optic (top row) and nano-photocurrent (bottom row) images acquired with incident light frequency ω=890 cm$^{-1}$.*

We first acquired near-field images using $\omega$=890 cm$^{-1}$ illumination. As mentioned in the main text, we collect the tip-scattered optical signal (nano-optic or NF) and photocurrent driven by the tip-assisted nanoscale temperature profile (nano-photocurrent or PC). To minimize the far-field background, the optical signal is demodulated at the 4$^{th}$ harmonic, and the photocurrent signal is demodulated at the 2$^{nd}$ harmonic of the tip-tapping frequency [1]. In Fig. S1, we demonstrate the doping-dependent NF and PC mapping of the graphene sample at 20 K. At below $\nu = 4$, strong magnetoplasmon fringes can be seen in the PC images.

To investigate the cause of the dramatic change of NF and PC contrast near and away from CNP, we perform repeated line scans across both regions 1 and 2 on the MLG while continuously tuning the back gate voltage. The line scans in **Fig. S2** are performed at 20 K and B=7 T using tunable monochromatic light ranging from $\omega$=856 cm$^{-1}$ to 925 cm$^{-1}$. Peak-like features clearly develop in the NF response within $\nu$=±6 and evolves dramatically with frequency, suggesting a small energy splitting of the 0$^{th}$ Landau level. The doping factor $\nu$ are labeled by horizontal dashed lines at each full filling of LLs at $\nu$=-2+4n (n∈Z) with $\Delta\nu$=4 corresponding to $\Delta V_g$≈10 V. This $\Delta\nu$=4 interval also coincides well with the ladder-like PC response in Fig. 2b, as we will address later.

To understand the resonance feature observed in NF responses (**Fig. S2a**), we first focus on NF experiments conducted with the lowest photon energy with a frequency $\omega$ at 856 cm$^{-1}$. Two peaks can be found near $\nu$=±2, leaving the CNP at a relatively lower signal level. The peak signal decays quickly to contrast-less before reaching $\nu$=±6, suggesting a strong 0$^{th}$ to ±1$^{st}$ LL excitation. At $\omega$=863 cm$^{-1}$, the two-peak feature maintains, whereas the signal at CNP raises. At $\omega$=871 cm$^{-1}$ and $\omega$=879 cm$^{-1}$, the signal at CNP is comparable to those of $\nu$=±2. At higher frequencies, e.g. $\omega$=925 cm$^{-1}$, the signal at CNP is more prominent than those at the electron side and hole side, forming a three-peak feature. The off-CNP peaks are now centered at $\nu$=±4. The evolution of these peak-like features versus frequency can be clearly seen from the vertical line-cuts within region 1 (indicated by the red arrows in **Fig. S2a**) at different frequencies (**Fig. S2c**). These line-cuts exhibit a gradual increase of signal at CNP and a transition from two-peak to three-peak structure with increasing photon energy.

The PC channel reveals more exotic real-space features around full filling ($\nu$=-2+4n (n∈Z)) and half-filling ($\nu$~4n (n∈Z)) of each LL. Similar to the NF response, the doping-dependent PC line scans in Fig. S1c have a strong signal strictly within the range -6<$\nu$<6, indicating the same origin from the LL transitions. The fringe-like patterns mentioned in manuscript **Fig. 1c** accompany this strong signal in **Fig. S2b** at all optical frequencies. The origin of the fringes is likely linked to the generation of the magnetoexcitons [2, 3, 4, 5]. At most doping levels above the ±1$^{st}$ LL, PC signal is 'short-ranged' and concentrates at two edges of the line scan (vertical 'rails' of the 'ladder'). The edge signal becomes more prominent at half-doped LLs ($\nu \sim 4n$ ($n \in \mathbb{Z}$)) than fully populated LLs. However, within a narrow range of full fillings ($\nu = -2 + 4n$ ($n \in \mathbb{Z}$)), the PC signal becomes 'long-ranged' and reoccupies the entire sample area, forming the 'rungs' of the 'ladder'. This clear bulk-edge correspondence can be understood within the clean limit of quantum hall physics, as we will discuss in later sections.

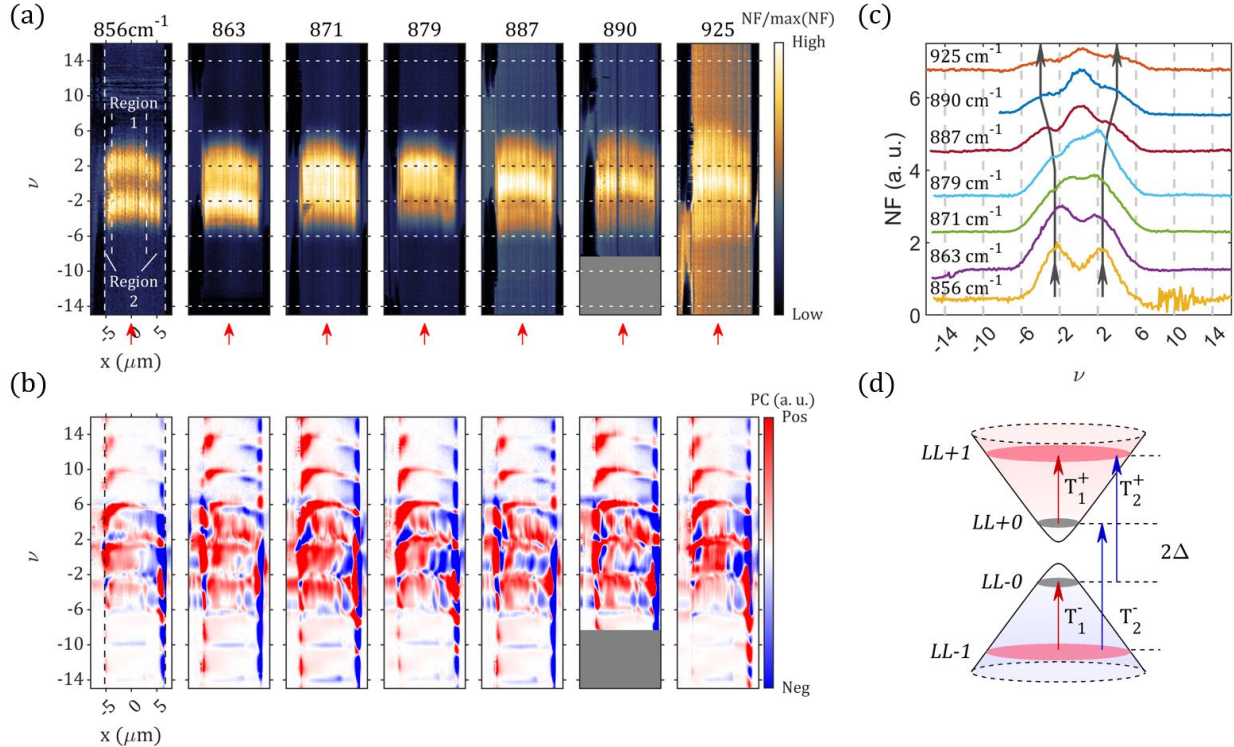

**Supplementary Figure 2|** *Doping- and frequency-dependent real-space profiles at T = 20 K and B = 7 T. (a) The doping dependences of the NF signal show multiple-resonance structures near CNP. When changing the optical frequency, the resonance structure exhibits a systematic transition from two-peak to three-peak feature. (b) The PC signal taken simultaneously with the NF signal in panel (a) shows a clear ladder-like structure. Within $-6<\nu<6$, the PC signal is strong due to magnetoexcitons. At higher doping levels, the PC signal periodically localizes near the edge of the sample and extends to the entire sample. (c) Gate-dependent NF signal evolution from low frequency (856 $cm^{-1}$) to high frequency (925 $cm^{-1}$). The linecuts are chosen in region 1 of MLG marked by red arrows in panel (a). At low frequencies, the NF signal near CNP comprises of two peaks centered around $\nu = \pm 2$. At high frequencies, the doping dependence evolves into three-peak structure centered around $\nu = 0, \pm 4$. The transition of off-CNP peaks in NF signal is highlighted by black arrows. (d) The $0^{th}$ Landau Level of MLG at high magnetic fields is split into two finer levels 0+ and 0-. The transitions between $LL \pm 1$ and $LL0\pm$ are denoted as $T_1^{\pm}$ for lower energy transitions and $T_2^{\pm}$ for higher energy transitions. These transitions contribute to the multiple peaks observed in the gate dependence near CNP.*

  The strong photon energy dependence of the NF resonances and PC responses within $-6<\nu<6$ strongly imply a finer structure in the optical transitions between the $\pm 1^{st}$ and the $0^{th}$ LL. The LLs of MLG exhibit 2-fold valley and 2-fold spin degeneracies. At high magnetic fields, the valley degeneracy is lifted [6, 7, 8], resulting in a split DOS of the $0^{th}$ LL near CNP with a gap of $2\Delta$. The optical transitions between the $0^{th}$ and the $\pm 1^{st}$ LL, therefore, involve four different transitions $T_1^{\pm}$ and $T_2^{\pm}$, as illustrated in **Fig. S2d**. The lower index 1 and 2 refers to the lower energy (intra-band) and higher energy (inter-band) transitions, respectively. The upper index + and – refers to the difference of angular momentum between the initial and final states. The transition energies can be written as $\omega = \sqrt{\Delta^2 + 2\hbar e v_F^2 B} \pm \Delta$. Using joint density of state analysis, we find the NF signals can be well explained by the four transitions involved with the 0$^{th}$ LL splitting. The two-peak feature corresponds to transitions $T_1^\pm$ at low photon energies while the three-peak feature corresponds to transitions $T_2^\pm$ at a higher photon energy (**Fig. S2d**). We note that the energy splitting occurs at slightly different gate voltages for regions 1 and 2 but nevertheless yields a similar gap size 2Δ, ruling out a strong impact from the substrate.

It is also worth noting that the strengths of the two off-CNP peaks are not symmetric with respect to the CNP. Before the peak structure transition (856 cm$^{-1}$ and 863 cm$^{-1}$), the peak at the hole side ($\nu$<0) is higher than the one at the electron side ($\nu$>0). After the transition ($\omega \geq$871 cm-1), the relative strengths of the off-CNP peaks reverse. This imbalance indicates transitions involving the -0$^{th}$ LL are stronger than those involving the +0$^{th}$ LL, which can be attributed to an inequality of the DOS in the -0$^{th}$ and +0$^{th}$ LL or the scattering rate of the transitions.

We can use joint density of state (JDOS) $\rho$ to understand the behavior of doping dependence. The JDOS between the initial (I) and final (F) state is $\rho \sim \sum g_I g_F f_I (1 - f_F)$, where $g$ represents the degeneracy and $f$ denotes the Fermi distribution. At CNP and zero temperature, LLs at the same side of CNP are either full or empty. As a result, the lower energy transitions are blocked $\rho_{T_1} \sim 0$, and the higher energy transitions have the maximized JDOS $\rho_{T_2} \sim 16$. With low energy photons resonating with $T_1$, no transition contributes to the JDOS at CNP ($\rho \sim 0$). For $T_1^-$, the transition has maximized JDOS $\rho \sim 8$ when the -1st LL is fully filled while the -0th LL is empty, therefore, signal due to $T_1^-$ is peaked at $\nu \sim -2$. By the same logic, signal due to $T_1^+$ is peaked at $\nu \sim 2$. This analysis explains the two-peak feature at $\omega \leq$863 cm-1. With higher photon energies resonating with $T_2$, both T2s contribute to the JDOS at CNP ($\rho \sim 16$). The JDOS of $T_2^-$ is maximized to $\rho \sim 8$ when the -1st LL is fully filled ($\nu \sim -4$), but only starts to decrease when the +0th LL starts to be filled ($\nu \sim 0$) and decreases to $\rho \sim 0$ when the +0th LL is fully filled ($\nu \sim 2$). Similarly, The JDOS of $T_2^+$ starts to rise at $\nu \sim -2$, maximizes between 0<$\nu$<4 and decreases to zero at $\nu \sim 6$. The overlapping between $T_2^\pm$ JDOS explains the three-peak feature and the doping span of signals observed at $\omega \geq$890 cm-1. At finite temperature (20 K), the small band gap 2Δ between the ±0$^{th}$ LL smear out the behavior near CNP due to Fermi-Dirac distribution. At the intermediate photon energies, both $T_1$ and $T_2$ can be excited due to a finite width of the LL energy distribution. At a specific frequency, $T_1$ and $T_2$ are excited with equal strength, giving rise to a plateau behavior of the optical signal around the CNP. This explains the result of the transition point from two-peak to three-peak feature at $\omega$=871 cm-1 and 879 cm-1. From the data, we can determine that the gap size is 2Δ≈30 cm-1 and the transition scattering rate is on the same scale Γ=30 cm-1. The transition frequency is $\omega = \sqrt{\Delta^2 + 2\hbar e v_F^2 B}$=875cm-1, giving rise to $v_F$=1.130×10$^6$ m/s. In summary, the analysis based on JDOS explains well the line shape of doping dependence at all frequencies.

In Region 1 of the device, we patterned SiO$_2$ layer of the Si wafer to induce artificial super-potential [9]. The pattern has hexagonal structure with lattice constant 50 nm. The vertex of the pattern is emptied SiO$_2$ spot with 30 nm diameter and 50~60 nm depth. The purpose of the pattern is to manually induce a splitting of the valley of degrees of freedom. However, without a top gate and with a 10 nm bottom hBN, the super-potential at low gating levels is not ideal. As a result, we observed nearly identical behavior inside and outside the patterned region. The only observable effect is a small offset in the CNP and a slight reduction in the doping efficiency.

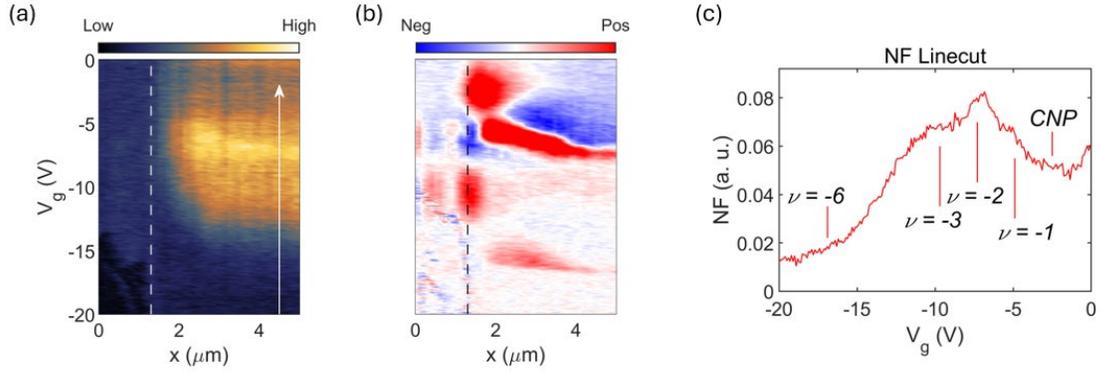

*Supplementary Figure 3| Near-field and photocurrent data at T=8 K and B=7 T. The near-field data is acquired with a photon frequency ω=857 cm-1. The illumination power in panel (a) and (b) is 200μW. (a) and (b) Similar to the measurement at 20 K, we keep scanning one line (the 5K measurement indicated in Fig. 1b) on the sample to acquire the doping dependence. The left edge of the sample is indicated by dashed lines. (c) The two-point transport data is comparable to the NF and PC data in panel (a) and (b). The conductance at the plateau slightly deviates from integer multiples of conductance quantum due to the contact resistance.*

We repeated the experiments in Fig. S2 at T=8 K. In **Fig. S3**, we collected data using optical frequency ω=856 cm-1, close to the $T_1$ transition. At this frequency, the doping-dependent optical response features two peaks on both sides of the CNP. The peak on the electron-doped side hosts finer structures. We take a line cut in **Fig. S3a** along the back gate voltages (white arrow) and show the result in **Fig. S3c**. These finer structures align well with the integer fillings of graphene at B=7 T. These features are likely induced by the observation of the further splitting of the -$0^{th}$ LL or the -$1^{st}$ LL at low temperatures.

## Section 2: Near-field signal of $0^{th}$ → $1^{st}$ LL transition probed by circular polarized light

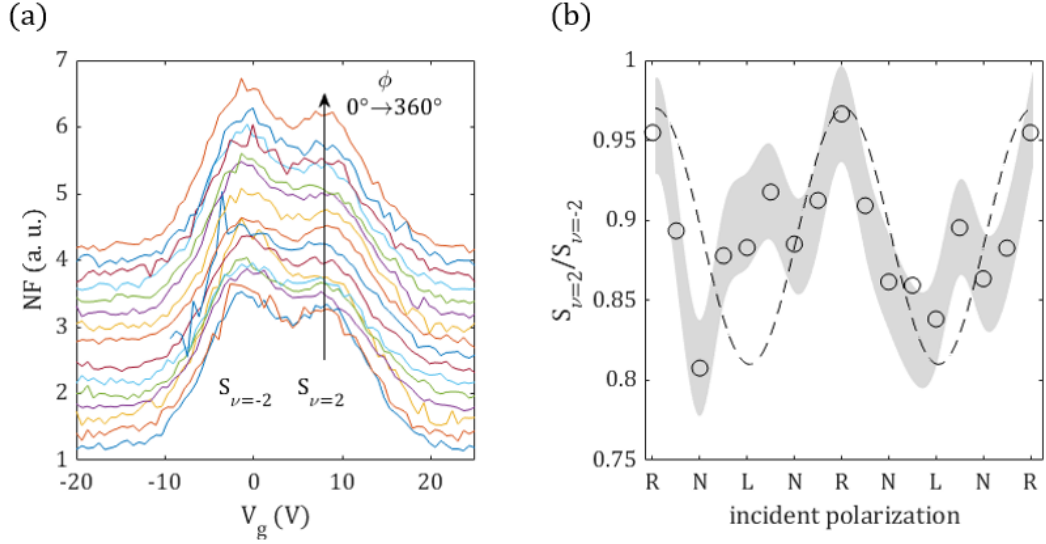

*Supplementary Figure 4| Nano-optical signal acquired using Circular polarized light. (a) We use a quarter waveplate to modulate the polarization of the incident beam. We mark the relative angle between the fast axis of the waveplate and the polarization of the incident beam. (b) We integrate the signal near $\nu=\pm2$ and calculate the ratio between two signals. According to the waveplate angle $\phi$, we mark the polarization of the incident beam. The notation R, N and L represent right-hand circular, linear and left-hand circularly polarized light.*

      In scattering-type near-field experiments, verifying or applying the selection rule faces several difficulties. First, the near-field tip, elongating in the direction perpendicular to the sample surface, highly favors a p-polarized incident light and contributes mainly to a p-polarized scattering field [10, 11, 12, 13]. Second, the near-field tip strongly confines the light field to its ~10 nm apex, contributing to a non-negligible momentum in the probe field. The small but finite momentum renders a smearing of the selection rule. Therefore, observing the polarization-dependent effects or carrying out polarization-dependent analyses, especially using circular polarization, is usually very challenging.

      The optical selection rule of the quantum Hall states of graphene suggests circularly polarized light (CPL) would selectively excite electron transition with raising or lowering quantum indices (for example -$1^{st}$ → -$0^{th}$ for the right-handed CPL and the +$0^{th}$ → $1^{st}$ for the left-handed CPL) [5]. When we illuminate graphene with low frequency light ($\omega$=856 cm$^{-1}$), the -$1^{st}$ → -$0^{th}$ and the +$0^{th}$ → $1^{st}$ LL transitions contribute to two separate peaks near the CNP. Since the optical selection indicates a 100% suppression/selection ratio of two peaks when illuminated with CPL, this provides a unique chance to verify the polarization sensitivity of near-field probes.

      We use a quarter waveplate to modulate the polarization of the incident beam and collect the power of the scattered signal without further polarization filtering. In **Fig. S4a**, we present doping-dependent signals while the quarter waveplate is setting at different angles. As mentioned in the manuscript and analyzed previously in the supplementary, the signal located at $\nu$=2 (-2) represents transition from +$0^{th}$ → $1^{st}$ (-$1^{st}$ → -$0^{th}$) LL. When tuning the waveplate, we didn't observe a 100% dependence of PC signal to the angle. Instead, we find a 20% variation of the contrast when the incident beam is alternatively tuned between right-handed, linear, and left-handed circularly polarized light. In **Fig. S4b**, we integrate and normalize the $\nu$=2 signal on the $\nu$=-2 signal to demonstrate the polarization dependence and compare the collected contrast (empty dots) to a standard sinusoidal curve (dashed line). Despite observable deviation,

we conclude that the near-field signal is still sensitive to the CPL. The suppression ratio of the acquired near-field signal is 10%.

## Section 3: Photocurrent Landau Fan

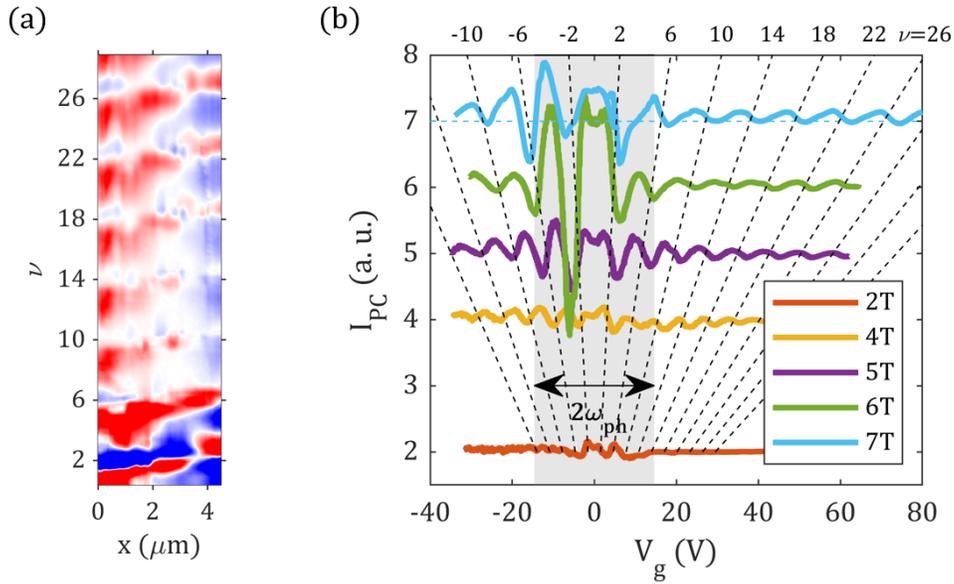

*Supplementary Figure 5|(a) Nano-photocurrent signal behaviors at high gate and high field (T = 20 K, B = 7 T). (b) Gate-dependent nano-photocurrent signal collected at various magnetic fields. We extract the signal on the left edge of the sample. The dashed lines are full-filling points ν=-2+4N (N∈Z). The grey region marks the doping range where inter-band transition can be excited in field-free graphene. This region also coincides with the -1$^{st}$ to the 1$^{st}$ LL of graphene at B=6~7T.*

The photocurrent signal (PC) in the quantum Hall regime is highly sensitive to defects and inhomogeneities in carrier density. Therefore, samples with different qualities demonstrate slightly different results. In the main text, we present the result acquired on a clean sample. Here in **Fig. S5a**, we present results acquired on a relatively dirtier sample at 7T. All defects in the sample produce a tiny PN junction which contributes to an observable signal. Despite the defects, the edge/bulk correspondence discussed in the manuscript is still visible in the data. We also note that the signal on the left (right) edge maintains predominantly positive (negative) sign. The behaviors persist to high gate levels. At high gates, due to the close energy spacing between neighboring Landau levels, the edge/bulk correspondence becomes less well-defined.

In **Fig. S5b**, we summarized the edge PC signal at various magnetic fields. The quantum oscillations in the edge PC signal are combined into a Landau fan. According to the carrier density, we mark the doping level ν=-2+4N(N∈Z) using dashed lines.

## Section 4: Magneto-plasmon at high magnetic field

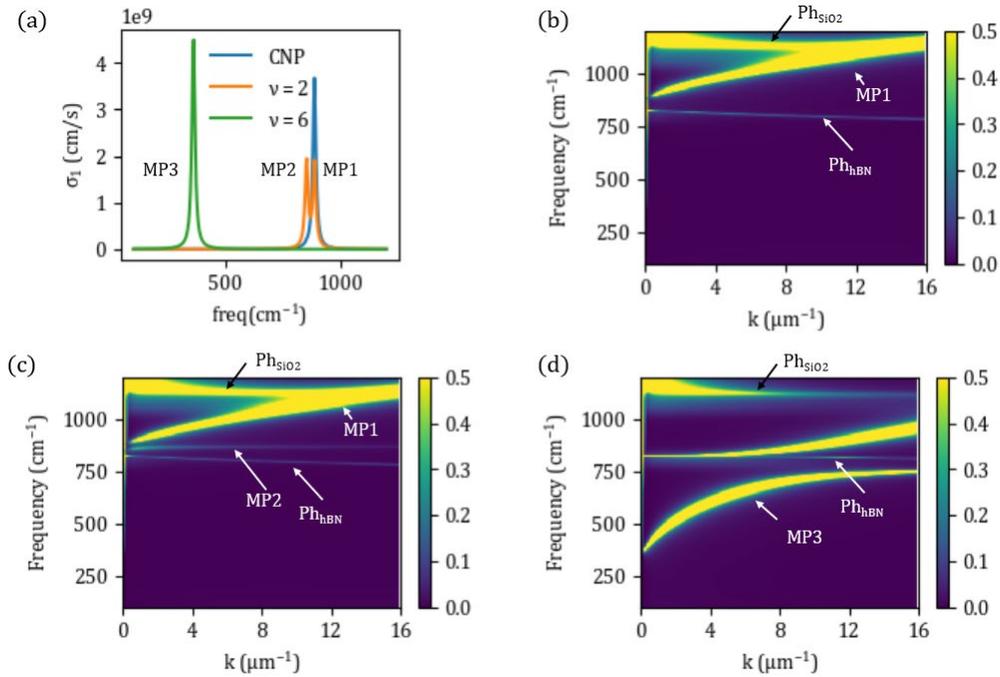

*Supplementary Figure 6| Optical response of graphene at high magnetic field and the corresponding magneto-plasmon. (a) The graphene optical conductivity calculated at T = 5 K, B = 7 T, G = 4 cm$^{-1}$ with chemical potential at CNP, v~2, and v~6. The three resonances are LL transitions from -0$^{th}$ →+1$^{st}$ (MP1, also -1$^{st}$ →+0$^{th}$ ), +0$^{th}$ →+1$^{st}$ (MP2, also -1$^{st}$ →-0$^{th}$ ) and -1$^{st}$ →+2$^{nd}$ (MP3, also -2$^{nd}$ →+1$^{st}$ ) (b)-(d), $r_p$ plots of the magneto-plasmon dispersion at CNP (b), v=2 (c), and v=6 (d).*

We follow Ref. [4] to calculate the optical conductivity of the graphene at high magnetic fields. We assume a finite gap opening at B = 7 T, 2Δ=30 cm$^{-1}$, $v_F$=1.12×10$^6$ m/s, T=5 K. To highlight the magneto-plasmon mode dispersion, the scattering rate is set to G=4 cm$^{-1}$. We summarized the optical conductivity at various doping levels in **Fig. S6a**. At CNP, due to the low temperature, the -0$^{th}$ LL is fully populated, whereas the +0$^{th}$ LL is empty. Only the $T_2$ modes (higher energy transition) can be triggered. At v=2, both -0$^{th}$ and +0$^{th}$ LLs are fully populated, contributing to two peaks ($T_1$ and $T_2$) in the optical conductivity. After fully filling the 1$^{st}$ LL (v=6), the 0$^{th}$ → 1$^{st}$ transition is blocked. The main optical response corresponding to the 1$^{st}$ to the 2$^{nd}$ LL transition is shifted to much lower frequency.

Each of these transitions contributes to a branch of magneto-plasmon dispersion in the momentum dependent reflectivity map (Imaginary part of r$_p$). We simulated the reflectivity as a function of frequency and momentum of a layered stack comprising of graphene, 10 nm hBN and SiO$_2$/Si substrate. The result is shown in **Fig. S6b~d**. Two modes located at ω=800 cm$^{-1}$ and 1200 cm$^{-1}$ originate from the phonons of the substrate material. The magneto-plasmon modes are typically dispersive, as shown in **Fig. S6b** and **S6d**. In **Fig. S6c**, due to the close spacing between the T$_1$ and T$_2$ mode, their response gives rise to a dispersive high energy mode and a low-lying flat mode.

## Section 5: Electron cooling time

We follow Bistritzer and Macdonald's [14] discussion on the time scale of the electron cooling to the lattice in graphene. Because the electron cooling model is not yet available, we approximate the cooling time scale in the high field scenario to field-free graphene with a similar carrier density. In the case of high doping, the time scale of a field-free graphene with a chemical potential $\mu$ can be calculated using the formula

$$\tau_1 \approx \frac{T_{el}}{0.133 D^2 n^{\frac{3}{2}}} \text{ ns}$$

Here the graphene deformation potential D is assumed to be 20 ([eV]), n ([$10^{12}$ cm$^{-2}$]) is the carrier density and $T_{el}$ is electronic temperature measured in meV. The carrier density of a half-filling LL is $n = 2\frac{eB}{h}$=0.34 [$10^{12}$ cm$^{-2}$]. With the above equation, we estimate the hall-filling cooling time $\tau_{half}$~1.65 ns.

We next approximate the full-filling scenario to the case near CNP. The time scale is,

$$\tau_2 \approx \frac{848}{D^2 T_L^2} \mu s$$

Here, $T_L$ represents the lattice temperature measured in meV. In our experimental scenario ($T_L$=5 K ~ 20 K), the cooling time scale is in the range $\tau_{full}$=700 ns ~ 11 μs.

**Section 6: Power dependence of the photocurrent signal**

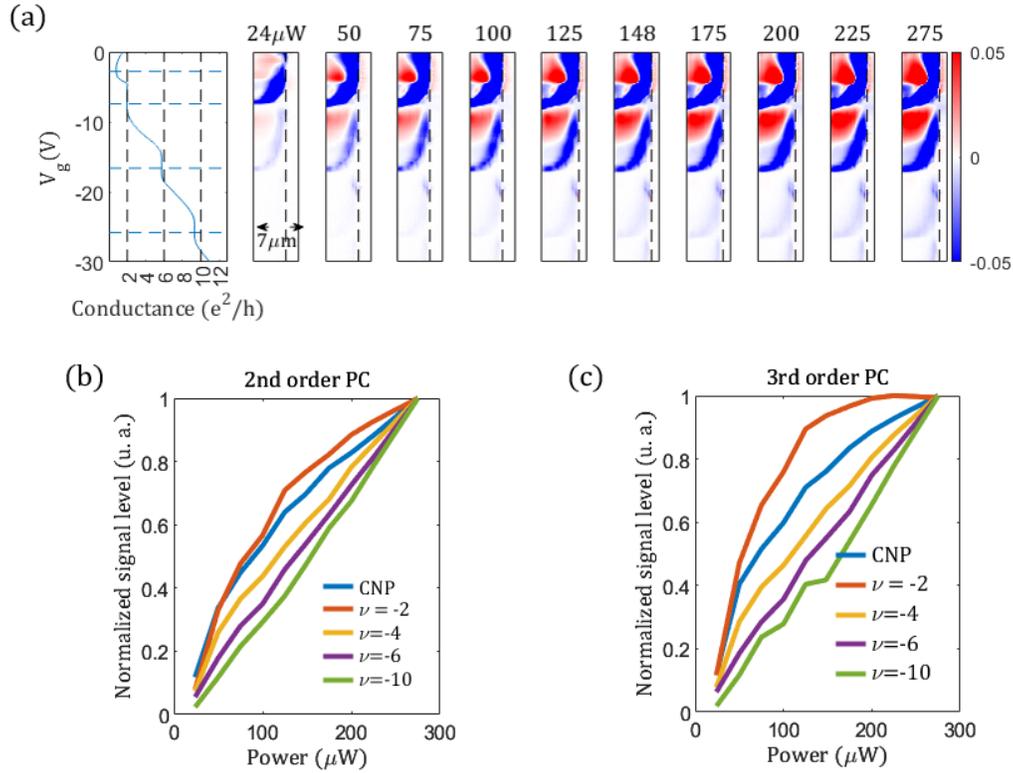

*Supplementary Figure 7 The power dependence of the photocurrent signal. (a) We repeat the photocurrent line-scans at a series of power from 20 µW to 275 µW. (b) The photocurrent signals are acquired at the 2nd and 3rd harmonics of the tip modulation. We take the signal in the interior of the sample and record the signal at ν=0, -2, -4, -6 and -10.*

At T = 5 K, we studied the illumination power dependence of the photocurrent signal. The thermal capacitance of graphene is expected to substantially decrease when the temperature decreases to below 10 K. Therefore, the focused beam on the sample may cause overheating of the entire sample and the local spot beneath the AFM tip. The heating effect is a function of illumination power and the absorption coefficient of the sample. The absorption of graphene to the illumination can be estimated based on the optical conductivity at the photon frequency. Therefore, the on-resonance LL transition would cause a substantial absorption. Beyond $\nu \sim 6$, the absorption is finite but low due to the optical conductivity contributed by the magneto-plasmon (the tail of the conductivity peak corresponding to the LL transition). In **Fig. S7a**, we repeated the gate-dependent photocurrent line-scan at an illumination power ranging from 24 µW to 275 µW. We observe gradually increasing signal levels at all doping levels. We note again that the photocurrent is driven by periodic heating at the frequency of the tip oscillation. We acquire the 2nd and 3rd harmonics of the demodulated photocurrent signal. The results are summarized in **Fig. S7b** and **Fig. S7c**. We observe clear deviation behaviors to the linear dependence of the $\nu$=0, -2 and -4 signals, whereas the $\nu$=-6 and -10 signals are linear. This observation indicates the strong absorption due to the LL transition substantially heat up the sample starting from below P=100 µW.

**Section 7: Photocurrent data acquired with $\omega$=0.8 THz**

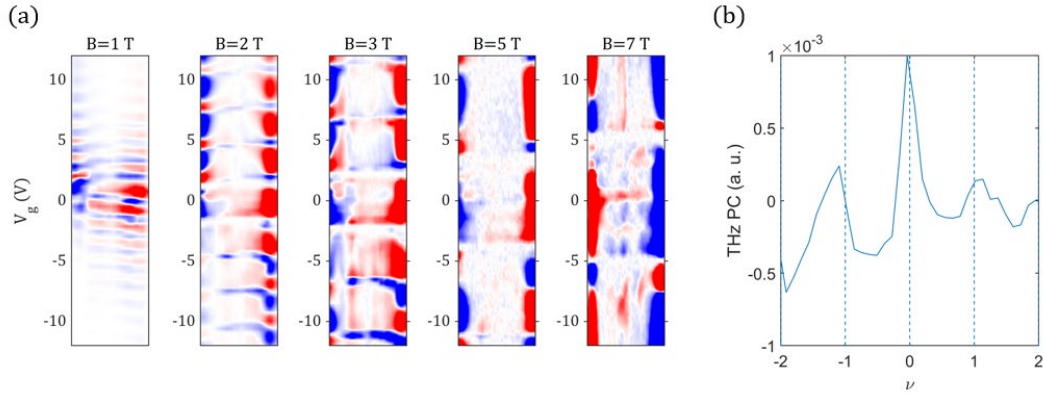

*Supplementary Figure 8|Doping dependence of photocurrent acquired at T=20 K using $\omega$=0.8 THz illumination. (a) Doping dependent line scans across free edges at various fields. (b) Doping dependent THz photocurrent signal at 7T.*

We acquired the nano-photocurrent on another high-quality graphene sample at terahertz frequency ($\omega$=0.8 THz). The terahertz source is a back-wave oscillator (Microtech Instruments, inc., QS1-900). The incident power to the near-field tip is 1mW.

The edge/bulk correspondence discussed in the manuscript can be reproduced in the THz data. At high field (B>1 T), the cyclotron frequency near CNP becomes much larger than the probe frequency $\omega$ =0.8 THz. Therefore, cyclotron resonance and magneto-plasmon are not likely responsible for the absorption of optical power. Magneto-edge-plasmon (MEP) is a branch of acoustic plasmon confined on the boundary of graphene [15, 16, 17]. It gives rise to finite response at THz frequencies along the graphene boundaries and around defects. We note that, due to the absence of cyclotron resonance and magneto-plasmon-induced strong signal, THz nano-photocurrent can further resolve more detailed behaviors of lifted degeneracies near the CNP. In **Fig. S8b**, we extract the gating dependence of THz nano-photocurrent in the bulk of graphene. We find that each integer filling point produces an observable bulk signal.

**Section 8: Shockly-Ramo Formalism applied in the quantum Hall regime**

Shockly-Ramo (SR) formalism has been successfully applied to explain photocurrent experiments on graphene, TMD and other materials which host strong photothermal effect [18]. With a local photocurrent generated $\boldsymbol{j}_{ph}$, the collected current signal in the circuit can be calculated with the help of an auxiliary field $\phi_{SR}$,

$$I_{PC} = A \int_S \boldsymbol{j}_{ph} \cdot \nabla \phi_{SR} \, d^2\boldsymbol{r}$$

The auxiliary field $\phi_{SR}$ is a weighting field on a sample with transposed conductivity $\sigma^T$, assuming current collecting contacts has a unit voltage drop between them. The pre-factor $A = R/(R + R_{ext})$ is the resistance ratio between the sample and the external circuit. We assume the local photocurrent in graphene is generated primarily due to the photothermal effect of the far-field illumination or the enhanced field on the tip apex. The local photocurrent can be described by the thermoelectric tensor $\alpha$,

$$\begin{bmatrix} j_{ph,x} \\ j_{ph,y} \end{bmatrix} = \begin{bmatrix} \alpha_{xx} & \alpha_{xy} \\ -\alpha_{xy} & \alpha_{xx} \end{bmatrix} \begin{bmatrix} \partial_x T \\ \partial_y T \end{bmatrix}$$

In the field-free case, thermoelectric effect is dominated by Seebeck effect $\alpha = \sigma S$. With magnetic field, the off-diagonal component becomes non-zero $\begin{bmatrix} \alpha_{xx} & \alpha_{xy} \\ -\alpha_{xy} & \alpha_{xx} \end{bmatrix} = \begin{bmatrix} \sigma_{xx} & \sigma_{xy} \\ -\sigma_{xy} & \sigma_{xx} \end{bmatrix} \begin{bmatrix} S & N \\ -N & S \end{bmatrix}$, due to the Nernst and Hall effect. For a uniform sample, we can isolate the terms contributed by the diagonal and off-diagonal terms,

$$I_{PC} = \int_{\partial S} \alpha_{xx} \delta T d\boldsymbol{n} \cdot \nabla \phi_{SR} - \int_{\partial S} \alpha_{xy} \delta T (d\boldsymbol{n} \times \nabla \phi_{SR})_z$$

We replace the scalar potential with the auxiliary electric field parallel or perpendicular to the free edge. The signal can be expressed as,

$$I_{PC} \propto \int_{\partial S} (\alpha_{xx} E_\perp^{SR} - \alpha_{xy} E_\parallel^{SR}) \delta T dl.$$

The calculation is valid for a uniform sample or for samples with multiple uniform regions. Under incident light illumination, the tip generates a nanoscale temperature profile $\delta T \propto \sigma_1(\omega)|E_{inc}|^2$. The integration takes place on the perimeter $\partial S$ of each sample region, which includes free edges, edges neighboring contacts, and internal boundaries forming PN junctions. The diagonal and off-diagonal thermoelectric coefficients, $\alpha_{xx}$ and $\alpha_{xy}$, can be both non-zero under magnetic field. The calculation involves a numerically solved auxiliary local field $E^{SR}$ with a 1V voltage drop between current-collecting contacts. The diagonal (off-diagonal) component $\alpha_{xx}$ ($\alpha_{xy}$) [19, 20, 21] contributes to the current signal $I_{PC}$ when the auxiliary field $E^{SR}$ has a perpendicular (parallel) component to the edge.

We first focus on the free edge of an ideally uniform sample. We consider the boundary elongates along y-axis and the sample is on the positive-x side. With the boundary condition of $j_x=0$ we have $j_x = \sigma_{xx} E_x^{SR} - \sigma_{xy} E_y^{SR} = 0$. The normal vector $\boldsymbol{n} = (-1,0)$. Therefore, $E_\perp^{SR} = -E_x^{SR}$ and $E_\parallel^{SR} = -E_y^{SR}$. Now,

$$I_{PC} \propto \int_{\partial S} (\alpha_{xx} \sigma_{xy} - \alpha_{xy} \sigma_{xx}) \frac{E_y^{SR}}{\sigma_{xx}} \delta T dl$$

For a long enough sample whose contacts can be considered far enough in the +y and -y direction, we can assume $E_y^{SR} \propto \sigma_{xx} \Delta V$, therefore

$$I_{PC} \propto (\alpha_{xx}\sigma_{xy} - \alpha_{xy}\sigma_{xx}) \int \delta T dl$$

We note that $\alpha_{xx} = \sigma_{xx}S - \sigma_{xy}N$ and $\alpha_{xy} = \sigma_{xx}N + \sigma_{xy}S$. Therefore,

$$I_{PC} \propto N \int \delta T dl$$

**Section 9: Tip induced signal without illumination**

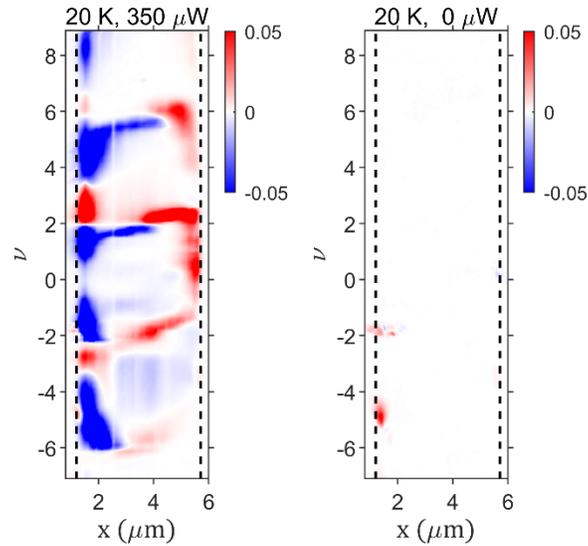

*Supplementary Figure 9|Tip modulated signal with illumination on and off. The injected photon frequency is 880 cm$^{-1}$.*

Without optical illumination, the AFM tip can also induce a weak current signal due to the single-electron charging effect [22, 23]. We specifically collected two doping-dependent line scans with light on and off in a row on another hBN encapsulated graphene device. On this device, with 350 µW illumination at the frequency $\omega$=880 cm-1, we observed nearly identical phenomena as other devices. Without illumination, the signal at most gating levels is suppressed. At a few doping levels, we observe repeatable weak signals that are one to two orders of magnitude smaller than the light-on case, especially at the sample edge. We conclude that the signal due to the single-electron charging effect is much weaker than the photo-induced current signal. However, the interplay between the tip-gated effect and the near-field photocurrent with very weak light illumination (<~10 µW) might be an interesting topic for future studies.

# Bibliography

[1]  A. J. Sternbach, J. Hinton, T. Slusar, A. S. McLeod, M. K. Liu, A. Frenzel, M. Wagner, R. Iraheta, F. Keilmann, A. Leitenstorfer and others, "Artifact free time resolved near-field spectroscopy," *Optics Express,* vol. 25, p. 28589–28611, 2017.

[2]  M. Dapolito, M. Tsuneto, W. Zheng, L. Wehmeier, S. Xu, X. Chen, J. Sun, Z. Du, Y. Shao, R. Jing and others, "Infrared nano-imaging of Dirac magnetoexcitons in graphene," *Nature Nanotechnology,* vol. 18, p. 1409–1415, 2023.

[3]  L. Wehmeier, S. Xu, R. A. Mayer, B. Vermilyea, M. Tsuneto, M. Dapolito, R. Pu, Z. Du, X. Chen, W. Zheng and others, "Nano-imaging of Landau-phonon polaritons in Dirac heterostructures," *arXiv preprint arXiv:2312.14093,* 2023.

[4]  V. P. Gusynin, S. G. Sharapov and J. P. Carbotte, "Magneto-optical conductivity in graphene," *Journal of Physics: Condensed Matter,* vol. 19, p. 026222, 2006.

[5]  I. O. Nedoliuk, S. Hu, A. K. Geim and A. B. Kuzmenko, "Colossal infrared and terahertz magneto-optical activity in a two-dimensional Dirac material," *Nature nanotechnology,* vol. 14, p. 756–761, 2019.

[6]  Y. Zhang, Z. Jiang, J. P. Small, M. S. Purewal, Y.-W. Tan, M. Fazlollahi, J. D. Chudow, <. f. J. A. Jaszczak, H. L. Stormer and P. Kim, "Landau-level splitting in graphene in high magnetic fields," *Physical review letters,* vol. 96, p. 136806, 2006.

[7]  E. A. Henriksen, P. Cadden-Zimansky, Z. Jiang, Z. Q. Li, L.-C. Tung, M. E. Schwartz, M. Takita, Y.-J. Wang, P. Kim and H. L. Stormer, "Interaction-induced shift of the cyclotron resonance of graphene using infrared spectroscopy," *Physical review letters,* vol. 104, p. 067404, 2010.

[8]  J. G. Checkelsky, L. Li and N. P. Ong, "Zero-energy state in graphene in a high magnetic field," *Physical review letters,* vol. 100, p. 206801, 2008.

[9]  J. Sun, S. A. Akbar Ghorashi, K. Watanabe, T. Taniguchi, F. Camino, J. Cano and X. Du, "Signature of correlated insulator in electric field controlled superlattice," *Nano Letters,* vol. 24, p. 13600–13606, 2024.

[10] T. Neuman, P. Alonso-González, A. Garcia-Etxarri, M. Schnell, R. Hillenbrand and J. Aizpurua, "Mapping the near fields of plasmonic nanoantennas by scattering-type scanning near-field optical microscopy," *Laser & Photonics Reviews,* vol. 9, p. 637–649, 2015.

[11] R. L. Olmon, P. M. Krenz, A. C. Jones, G. D. Boreman and M. B. Raschke, "Near-field imaging of optical antenna modes in the mid-infrared," *Optics express,* vol. 16, p. 20295–20305, 2008.

[12] J. D'Archangel, E. Tucker, E. Kinzel, E. A. Muller, H. A. Bechtel, M. C. Martin, M. B. Raschke and G. Boreman, "Near-and far-field spectroscopic imaging investigation of resonant square-loop infrared metasurfaces," *Optics Express,* vol. 21, p. 17150–17160, 2013.

[13] R. Ren, X. Chen and M. Liu, "High-efficiency scattering probe design for s-polarized near-field microscopy," *Applied Physics Express,* vol. 14, p. 022002, 2021.

[14] R. Bistritzer and A. H. MacDonald, "Electronic cooling in graphene," *Physical Review Letters,* vol. 102, p. 206410, 2009.

[15] I. Crassee, M. Orlita, M. Potemski, A. L. Walter, M. Ostler, T. Seyller, I. Gaponenko, J. Chen and A. B. Kuzmenko, "Intrinsic terahertz plasmons and magnetoplasmons in large scale monolayer graphene," *Nano letters,* vol. 12, p. 2470–2474, 2012.

[16] H. Yan, Z. Li, X. Li, W. Zhu, P. Avouris and F. Xia, "Infrared spectroscopy of tunable Dirac terahertz magneto-plasmons in graphene," *Nano letters,* vol. 12, p. 3766–3771, 2012.

[17] V. A. Volkov and S. A. Mikhailov, "Edge magnetoplasmons: low frequency weakly damped excitations in inhomogeneous two-dimensional electron systems," *Sov. Phys. JETP,* vol. 67, p. 1639–1653, 1988.

[18] J. C. W. Song, M. S. Rudner, C. M. Marcus and L. S. Levitov, "Hot carrier transport and photocurrent response in graphene," *Nano letters,* vol. 11, p. 4688–4692, 2011.

[19] J. G. Checkelsky and N. P. Ong, "Thermopower and Nernst effect in graphene in a magnetic field," *Physical Review B—Condensed Matter and Materials Physics,* vol. 80, p. 081413, 2009.

[20] Y. M. Zuev, W. Chang and P. Kim, "Thermoelectric and magnetothermoelectric transport measurements of graphene," *Physical review letters,* vol. 102, p. 096807, 2009.

[21] V. P. Gusynin and S. G. Sharapov, "Transport of Dirac quasiparticles in graphene: Hall and optical conductivities," *Physical Review B—Condensed Matter and Materials Physics,* vol. 73, p. 245411, 2006.

[22] M. T. Woodside and P. L. McEuen, "Scanned probe imaging of single-electron charge states in nanotube quantum dots," *Science,* vol. 296, p. 1098–1101, 2002.

[23] D. Halbertal, J. Cuppens, M. B. Shalom, L. Embon, N. Shadmi, Y. Anahory, H. R. Naren, J. Sarkar, A. Uri, Y. Ronen and others, "Nanoscale thermal imaging of dissipation in quantum systems," *Nature,* vol. 539, p. 407–410, 2016.